\documentclass[a4paper,fleqn,usenatbib]{mnras}

\usepackage{newtxtext,newtxmath}

\usepackage[T1]{fontenc}
\usepackage{ae,aecompl}

\usepackage{graphicx}	
\usepackage{amsmath}	
\usepackage{amssymb}	
\usepackage{bm}
\usepackage{epsfig}
\usepackage{threeparttable}
\usepackage[justification=centering]{caption}
\usepackage{times}
\usepackage[T1]{fontenc}
\usepackage{aecompl}
\usepackage{float}
\usepackage{multirow}

\title[The partial ionisation zone of heavy elements in F-stars]{The partial ionisation zone of heavy elements in F-stars: a study on how it correlates with rotation}

\author[Ana Brito \& Il\'idio Lopes]{
Ana Brito,$^{1,2}$\thanks{E-mail: ana.brito@tecnico.ulisboa.pt}
Il\'idio Lopes,$^{1}$
\\
$^{1}$Centro de Astrof\'{\i}sica e Gravita\c c\~ao  - CENTRA, Departamento de F\'{\i}sica, Instituto Superior T\'ecnico \\ IST, Universidade de Lisboa - UL, Av. Rovisco Pais 1, 1049-001 Lisboa, Portugal\\
$^{2}$Departamento de Matem\'atica, Instituto Superior de Gest\~ao\\ Av. Marechal Craveiro Lopes, 1700-284, Lisboa, Portugal\\
}

\date{Accepted XXX. Received YYY; in original form ZZZ}

\pubyear{2015}

\begin{document}
\label{firstpage}
\pagerange{\pageref{firstpage}--\pageref{lastpage}}
\maketitle

\begin{abstract}
	
We study the relation between the internal structures of 10 benchmark main-sequence F-stars and their rotational properties.
Stellar rotation of main-sequence F-type stars  can be characterised by two distinct rotational regimes. Early-type F-stars are usually rapid rotators with periods typically below 10 days, whereas later-type F-stars have longer rotation periods. 
Specifically, and since the two rotational regimes are tightly connected to the effective temperatures of the stars, we investigate in detail the characteristics of the partial ionisation zones in the outer convective envelopes of these stars, which in turn, depend on the internal temperature profiles.

Our study shows that the two rotational regimes might be distinguished by the relative locations of the partial ionisation region of heavy elements and the base of the convective zone. 
Since in all these stars is expected a dynamo-driven magnetic field where the shear layer between convective and radiative zones (tachocline) plays an important role, this result suggests that the magnetic field may be related to the combined properties of convection and ionisation.
\end{abstract}




\section{Introduction}\label{sec1}

\begin{figure*}
	\includegraphics[width=1.0\textwidth]{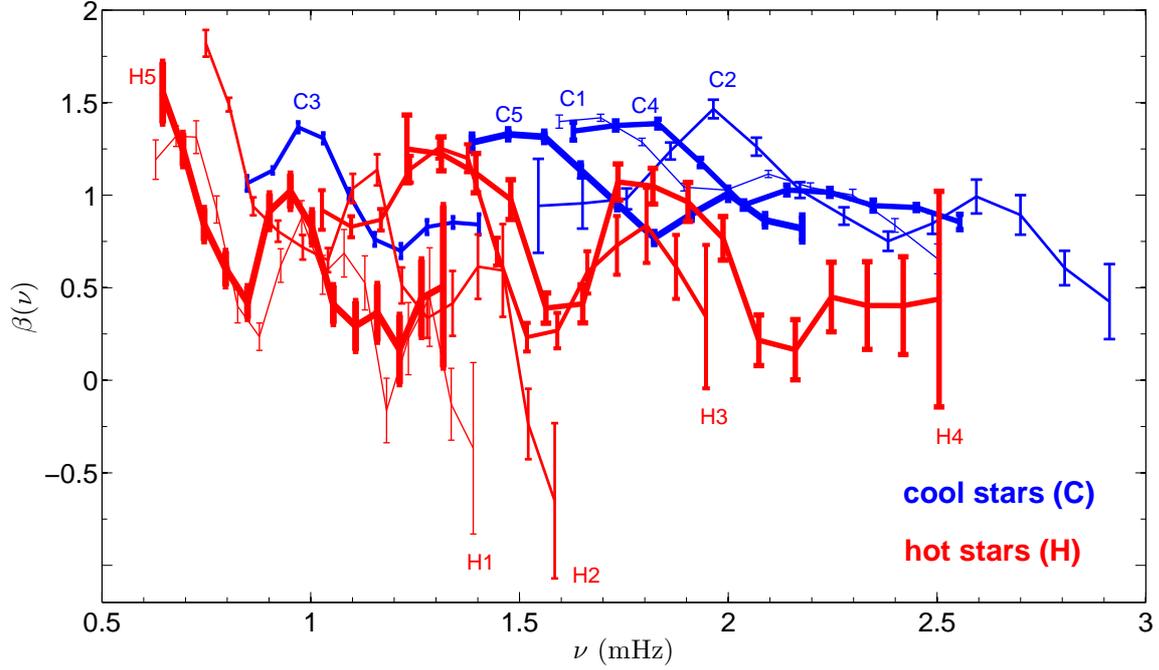}
	\caption{The seismic observable $\beta(\nu)$ computed for 10 main-sequence  F-stars in the case of oscillation frequency modes with degree $\ell=1$. Cooler stars are represented with blue colour and hotter stars are represented with red colour. Identification of the \emph{Kepler} stars: C1 (KIC 6116048), C2 (KIC 10454113), C3 (KIC 8228742), C4 (KIC 10963065), C5 (KIC 12009504), H1 (KIC 6679371), H2 (KIC 7103006), H3 (KIC 1435467), H4 (KIC 9206432), H5 (KIC 6508366). The numbering from 1 to 5 intents to distinguish the stars. Number 1 correspond to the thiner line whereas number 5 corresponds to the thicker line.}
	\label{fig1} 
\end{figure*}

Understanding how the microphysics of the stellar interiors influences and shapes the observable characteristics of stars is a challenging exercise. The evolution and structure of a star is conditioned by how the global stellar characteristics respond to the relevant microphysics processes, such as, for instance, radiative opacities, equations of state, diffusion and settling of chemical elements and also nuclear rates of energy production.
The present time, as the era of asteroseismic space-borne missions, is ideal to put constraints on the microphysics related processes occurring in the interiors of stars, from the exquisite observational data available from the missions CoRoT \citep{2007AIPC..895..201B, 2008CoAst.156...73M}, \emph{Kepler} \citep{2010Sci...327..977B, 2010ApJ...713L..79K}, the re-purposed K2 \emph{Kepler} mission \citep{2014PASP..126..398H}, and more recently from the TESS \citep{2014SPIE.9143E..20R} mission. In the future, this possibility will add to the data from the PLATO mission \citep{2014ExA....38..249R}.

In a recent work, using asteroseismology, \citet{2017ApJ...843...75B} unveiled a correlation, in the form of a power law, between the intensity of partial ionisation processes occurring in the outer envelopes of F-type stars and the surface rotation rate of these stars. 
Here, we explore the above cited ionisation--rotation relation to better understand the different internal structures of F-stars and  their relation to the distinct patterns of rotation on the main sequence.

\begin{table}
	\centering
	\large
	\begin{threeparttable}
		\caption{Observational parameters}
		\begin{tabular}{ c c c }          
			\hline
			\hline
			Star Id. & $T_{\text{eff}}$ (K) & [Fe/H] (dex)  \\
			\hline
			\hline
			C1 & 6033 $\pm$ 77    & -0.23 $\pm$ 0.1  \\
			C2 & 6177 $\pm$ 77    & -0.07 $\pm$ 0.1  \\
			C3 & 6122 $\pm$ 77    & -0.08 $\pm$ 0.1  \\
			C4 & 6140 $\pm$ 77    & -0.19 $\pm$ 0.1  \\
			C5 & 6179 $\pm$ 77    & -0.08 $\pm$ 0.1  \\
			\hline
			\hline
			H1 & 6479 $\pm$ 77    &  0.01 $\pm$ 0.1  \\
			H2 & 6344 $\pm$ 77    &  0.02 $\pm$ 0.1  \\
			H3 & 6326 $\pm$ 77    &  0.01 $\pm$ 0.1  \\
			H4 & 6538 $\pm$ 77    &  0.16 $\pm$ 0.1  \\
			H5 & 6331 $\pm$ 77    & -0.05 $\pm$ 0.1  \\
			\hline
			\hline
		\end{tabular}
		\label{table:1}
	\end{threeparttable}
\end{table}

\begin{figure*}
	\includegraphics[width=1.0\textwidth]{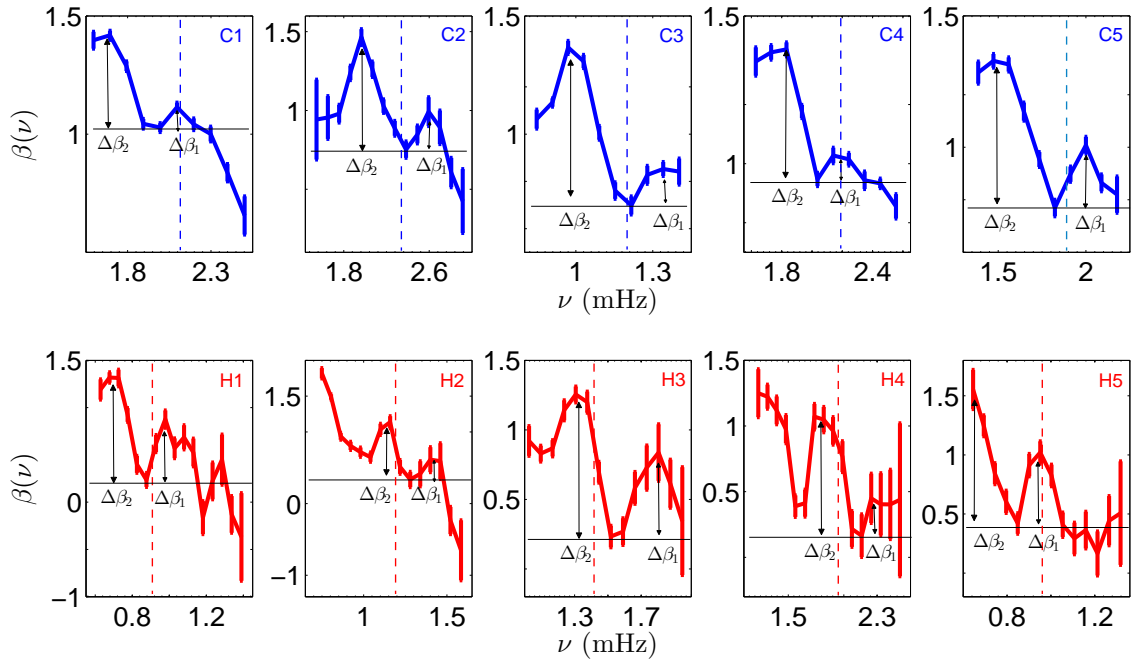}
	\caption{Schematic representation of the ionisation indexes $\Delta \beta_1$ and $\Delta \beta_2$ for all the stars in Figure \ref{fig1}. The vertical dashed lines signal the locations of the observational values of the frequency of maximum power, $\nu_{\text{max}}$, for each star. The indexes represent the amplitudes of the local maxima closer to $\nu_{\text{max}}$ \citep{2017ApJ...843...75B}.}
	\label{fig2} 
\end{figure*}

The observable data on stellar rotation provides valuable information to describe star formation and evolution. The behaviour of angular momentum evolution for low-mass stars is complex and depends on many interconnected factors. During the initial stages of star formation, the initial conditions both dynamical and chemical play a crucial role. In the pre-main sequence (PMS) phase of evolution of the stellar structure, phenomena like disk-locking, core-envelope decoupling and internal transport of angular momentum define the characteristics of the star when it reaches the zero-age main sequence \citep[e.g.,][]{2010ApJ...716.1269D, 2012ApJ...746...43R, 2014prpl.conf..433B}. 
At the moment when stars reach the zero age main sequence (ZAMS) they exhibit a large dispersion in the spin rates which is a consequence of the different physical processes occurring in the PMS evolution time. What happens next on the main sequence is strongly dependent on mass. Higher-mass stars ($\gtrsim 1.3  \, \, M_{\odot}$) keep the rotation rates with which they arrived at ZAMS, i.e., these stars conserve the angular momentum. In contrast, lower mass stars ($\lesssim 1.3  \, \, M_{\odot}$) begin to spin down and lose angular momentum. This spin down is itself very dependent on the mass of the stars with the less massive stars spinning down slower  than the more massive ones. By an age of approximately of 1 Gyr  all low mass stars have converged to a similar rotation rate \citep[e.g.,][]{2013A&A...556A..36G, 2015A&A...577A..98G}. 

The transition between the two rotational regimes on the main sequence -- high-mass stars that conserve angular momentum, and low mass stars that lose angular momentum -- can be considered steep in the sense that the mean values of the stars' projected rotational velocities suffer a break that occur over a small mass range. Specifically, as observed by \citet{1967ApJ...150..551K} for A, F, and G-type stars (using around 200 field stars) the steep variation in velocities from $\sim 20 \, \text{km} \, \text{s}^{-1}$ to $\sim 80 \, \text{km} \, \text{s}^{-1}$ occurs, respectively, between  $1.2-1.4 \, M_{\odot}$, which in turn corresponds to the vicinity of the spectral type F5 V. This transition is, consequently, known as the Kraft break. 
It is thought that the spin down and the consequent slow rotation of the lower mass stars ($\lesssim 1.3  \, \, M_{\odot}$) is due to angular momentum loss through magnetised stellar winds \citep[e.g.,][]{1967ApJ...148..217W}. These magnetised stellar winds act by exerting a braking torque on the star's rotation causing the spin down of the star \citep[e.g.,][]{1995A&A...294..469K, 2009IAUS..258..363I, 2015A&A...577A..27J}.
It is also thought that the extension of the convection zones in sun-like stars is directly related with the efficiency in generating the magnetic winds.
This is because stellar matter lost from the surface of the star should be kept in corotation with the star. This corotation, of the lost matter with the star, in turn, should be supported by a magnetic field.
Thus, it is argued that cooler stars, with deeper convective envelopes, are more effective in generating such a supporting magnetic field through the dynamo action \citep[e.g.,][]{1962AnAp...25...18S, 1968MNRAS.138..359M}.The theoretical parameterisations for the angular momentum loss are usually proportional to some power of the surface rotation rate \citep[e.g.,][]{1988ApJ...333..236K, 1997ApJ...480..303K}. For explaining the losses of angular momentum according to recent observational data on rotation several theoretical models are being proposed \citep[e.g.,][]{2015A&A...577A..98G, 2015A&A...577A..27J, 2017MNRAS.472.2590S}.

\begin{figure*}
	\includegraphics[width=0.49\textwidth]{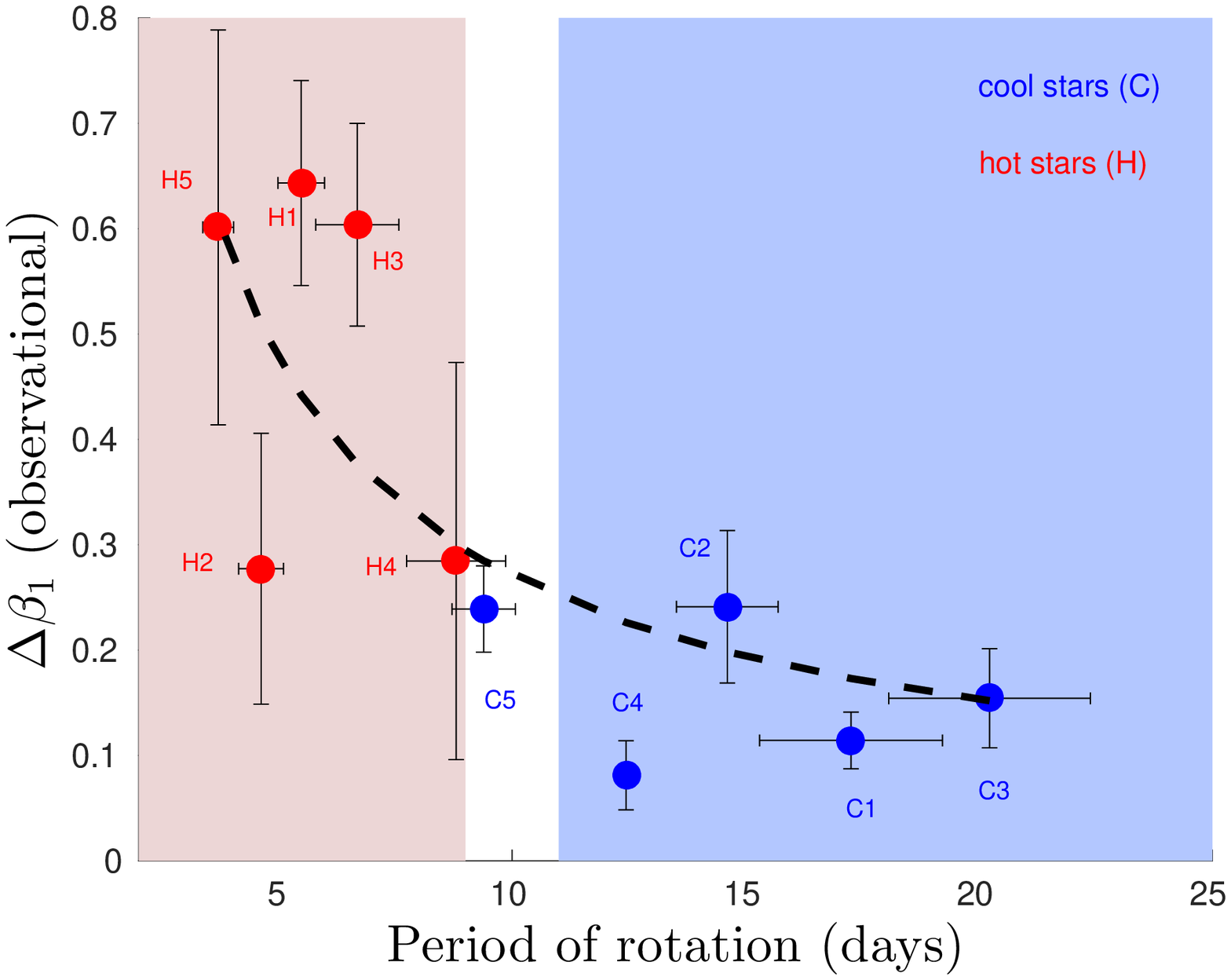}
	\includegraphics[width=0.49\textwidth]{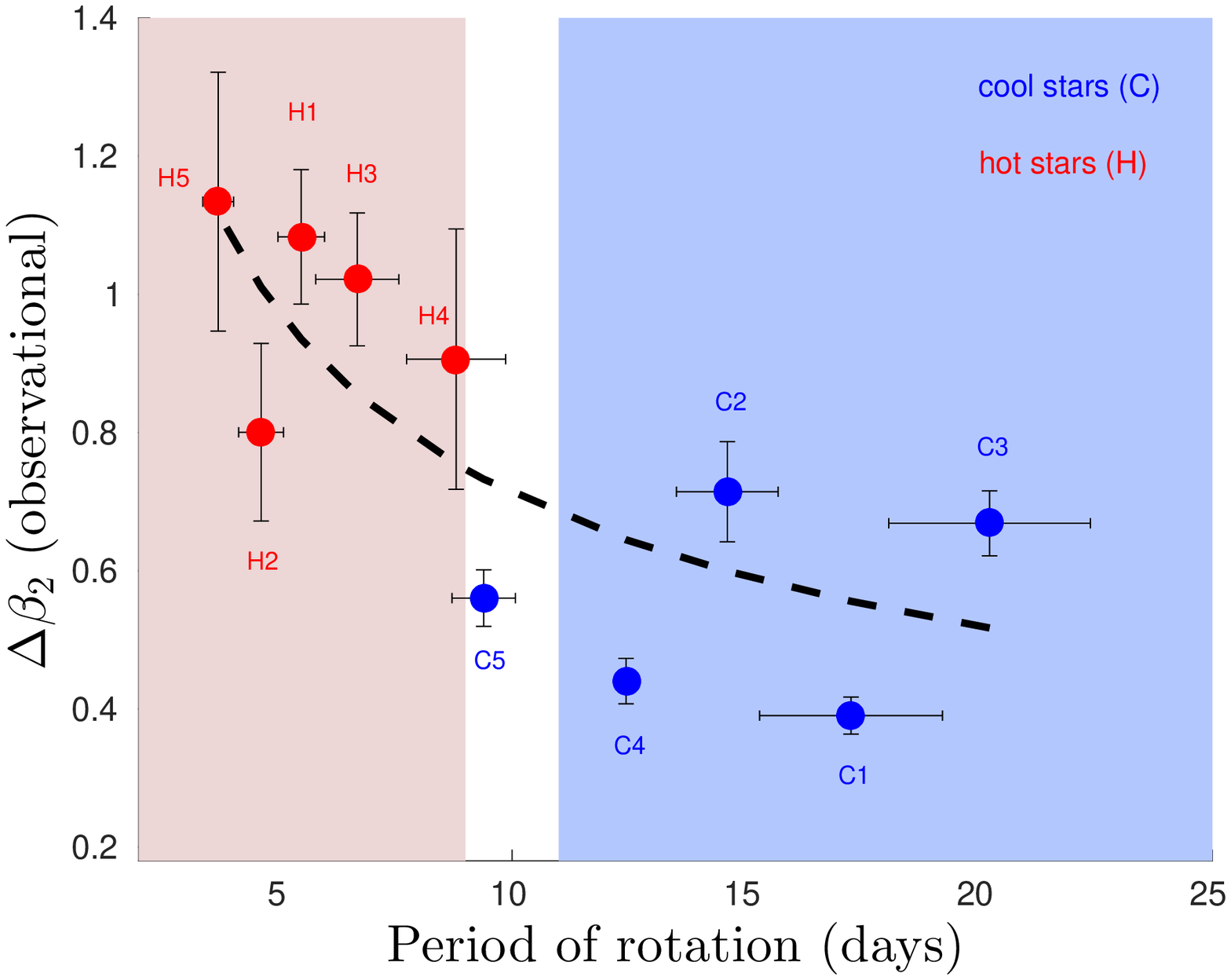}\\
	\includegraphics[width=0.49\textwidth]{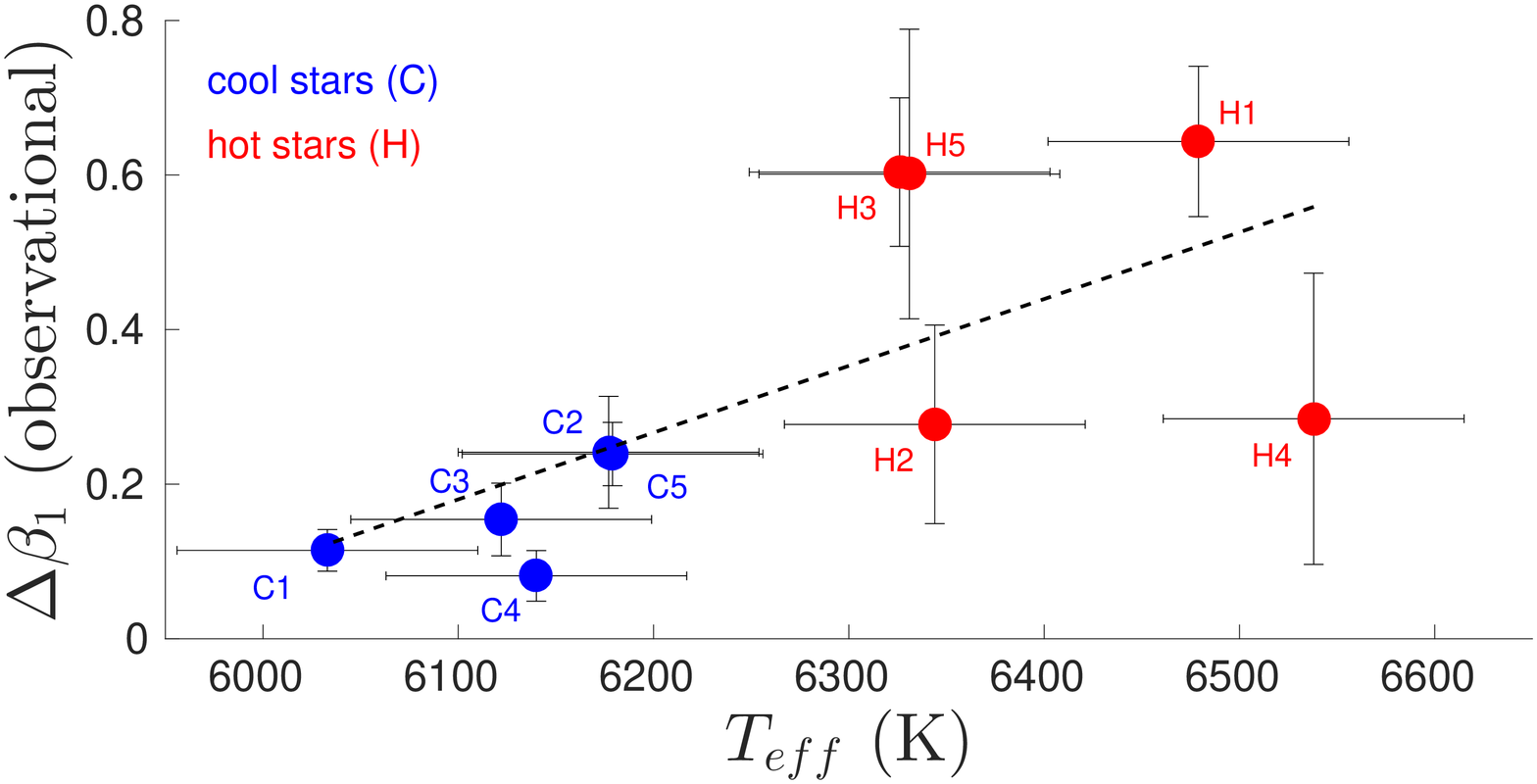}
	\includegraphics[width=0.49\textwidth]{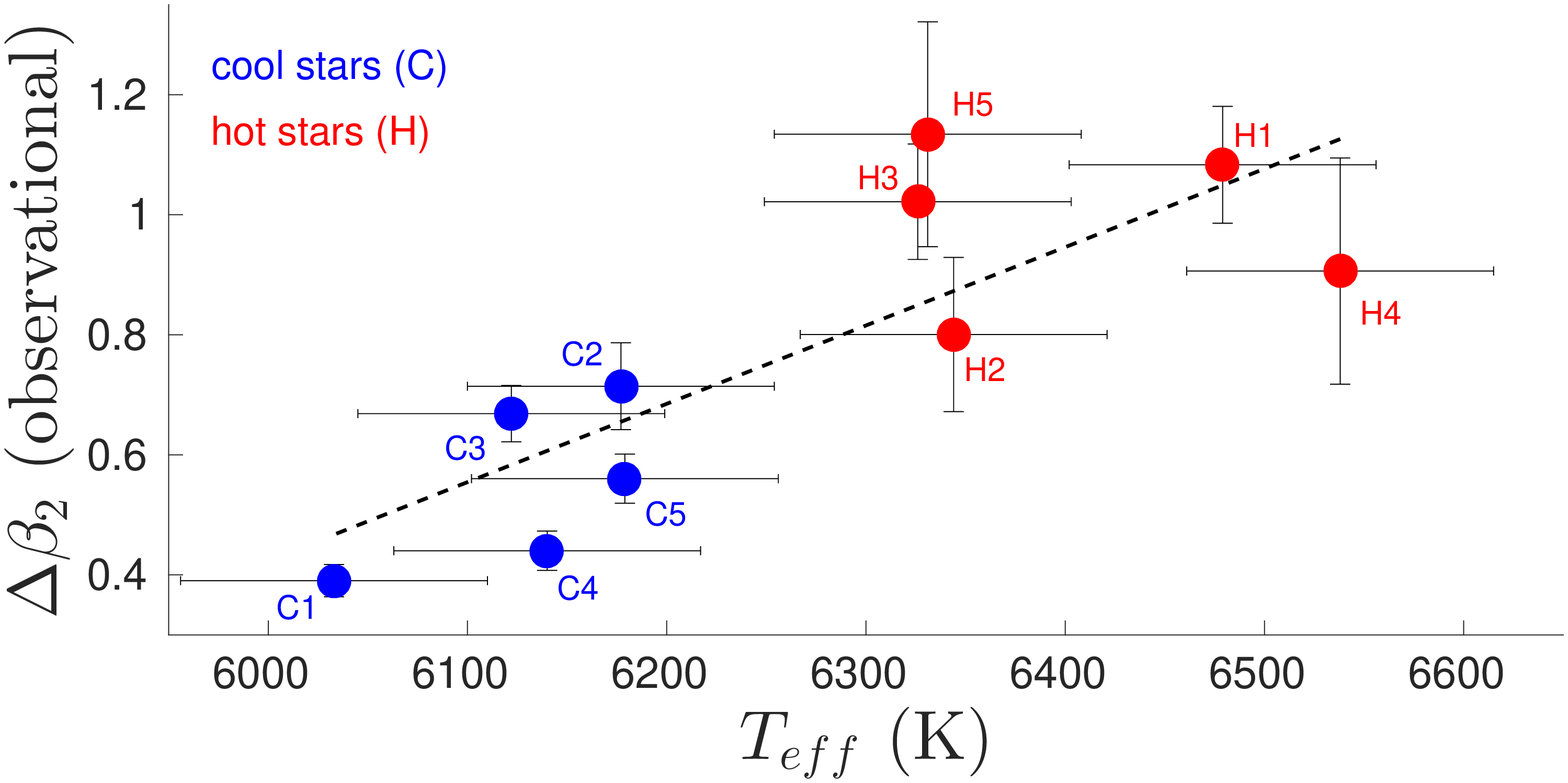}\\
	\includegraphics[width=0.49\textwidth]{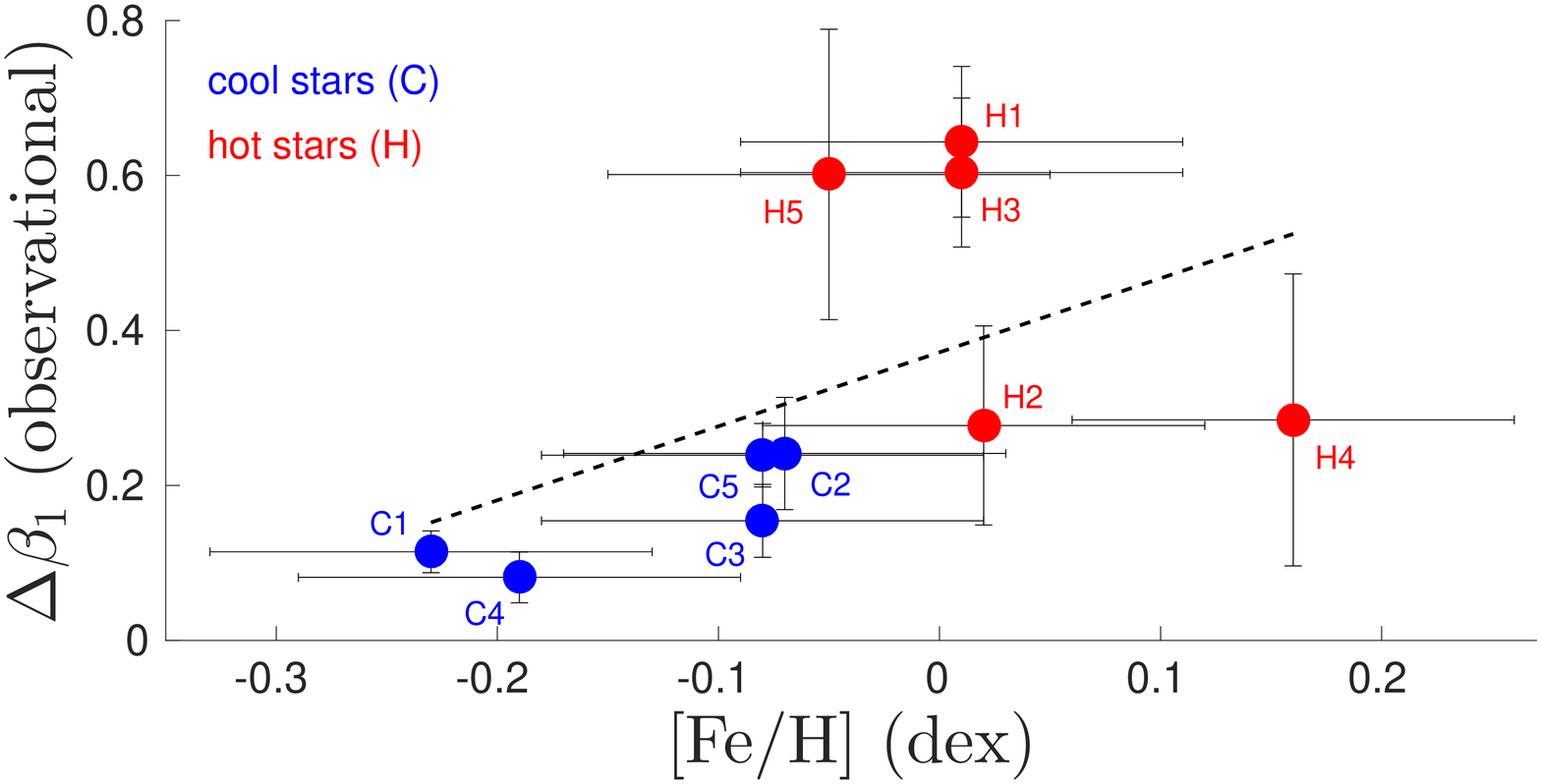}
	\includegraphics[width=0.49\textwidth]{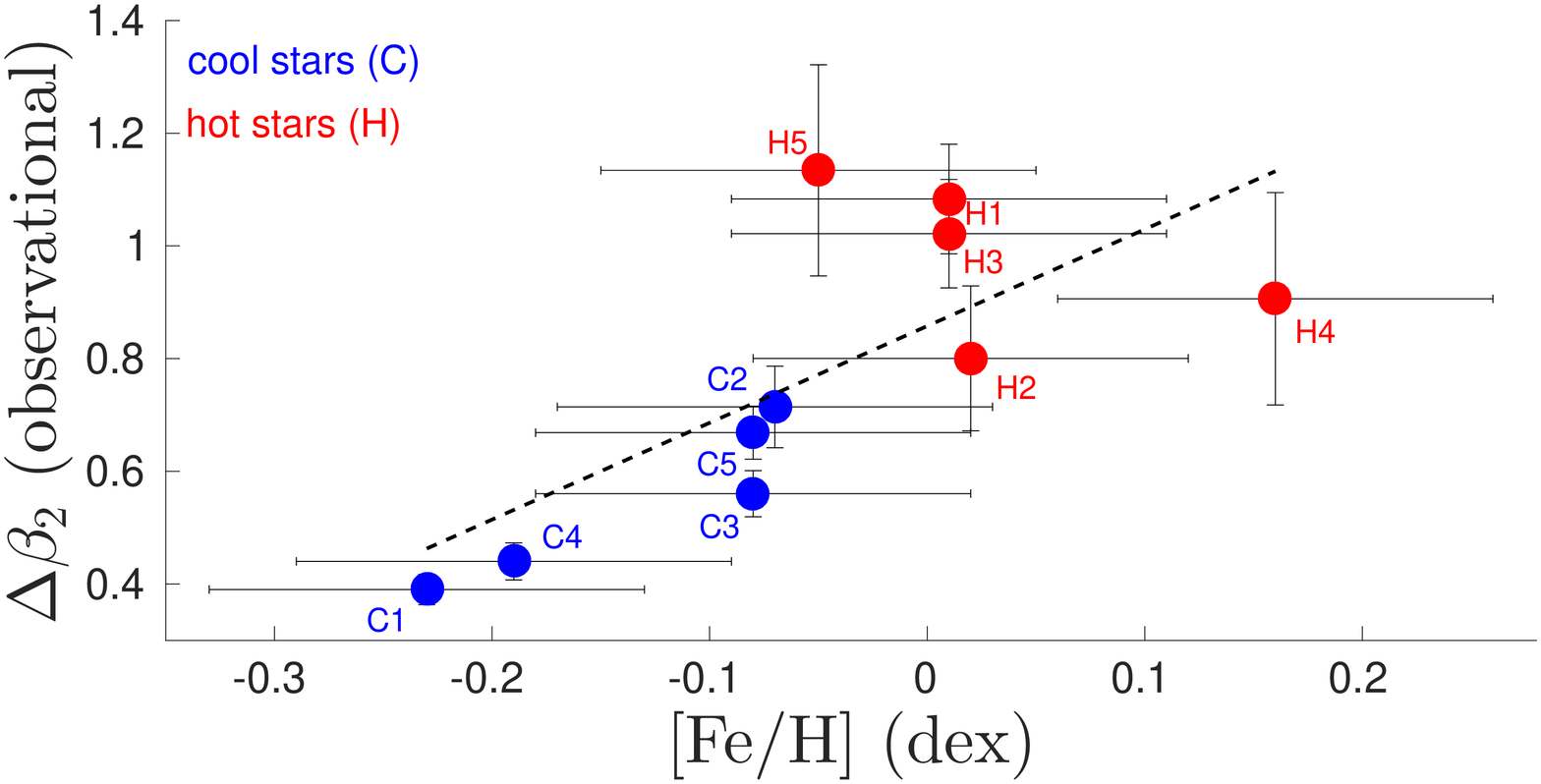}
	\caption{Upper panel: the ionisation indexes $\Delta \beta_1$ and $\Delta \beta_2$ plotted against the rotation period of the stars. The dashed line represents the result of a power law fit, $\Delta \beta_i = a \times P_{\text{rot}}^b$ to the data ($i=1,2$). Specifically,  $\Delta \beta_1 = 1.777 \times P_{\text{rot}}^{(-0.8176)}$ and  $\Delta \beta_2 = 2.023 \times P_{\text{rot}}^{(-0.4536)}$. The vertical white bar signalises the values of the rotation period around which occurs the transition between the two rotational regimes on the main sequence (the Kraft break). Rotation periods were taken from \citet{2015MNRAS.452.2654B, 2017A&A...598A..77K}. Central panel: the ionisation indexes $\Delta \beta_1$ and $\Delta \beta_2$ plotted against the observational effective temperature of stars. The dashed line represents the result of a linear fit, $\Delta \beta_i = a \times T_{\text{eff}} + b$ to the data ($i=1,2$). Specifically,  $\Delta \beta_1 = 8.643 \times 10^{-4} \, T_{\text{eff}} - 5.092 \, $ and $ \, \Delta \beta_2 = 1.305 \times 10^{-3} \, T_{\text{eff}} - 7.406$. Bottom panel: the ionisation indexes $\Delta \beta_1$ and $\Delta \beta_2$ plotted against the observational metallicity of stars. The dashed line represents the result of a linear fit, $\Delta \beta_i = a \times T_{\text{eff}} + b$ to the data ($i=1,2$). Specifically,  $\Delta \beta_1 = 0.955 \times T_{\text{eff}} + 0.3718$ and $\Delta \beta_2 = 1.717 \times T_{\text{eff}} + 0.8578$. Effective temperatures and metallicities were taken from \citet{2017ApJ...835..173S}.}
	\label{fig3}
\end{figure*}

In this paper we study a sample of ten benchmark Kepler F-stars that cover the interesting transitional region  near the Kraft break. The five cooler stars of the sample will be referred as cool stars and are located on the cool side of the Kraft break. The hotter five stars will be called hot stars and are located on the hot side of the Kraft break.  We show that the transition in the rotational regimes is correlated with a fundamental physical process occurring in stellar interiors -- the partial ionisation of chemical elements. 
Below the surface, and also below the well studied region of the second ionisation of helium \citep[e.g.,][]{1990LNP...367..283G, 1997ApJ...480..794L, 2004MNRAS.350..277B, 2007MNRAS.375..861H}, there is another important ionisation zone related to the ionisation of heavier elements, such as, carbon, nitrogen and oxygen \citep[e.g.,][]{2017MNRAS.466.2123B}. The possibility of using heavy-element ionisation to determine solar abundances was studied in the past by \citet{2009ASPC..416..203M}. 
The impact of this ionisation on the oscillations was recently shown to be comparable to the impact of the well known helium ionisation in hotter F-stars \citep{2018ApJ...853..183B}. Specifically, we show that 
there is a particular structural difference between late and early F-type stars, and that this difference is related to the combined locations of the ionisation zone of heavy elements and the base of the convective envelope.
Moreover, the transition in the rotation rates of these stars seems to be  correlated to this relative position of the base of the convective zone (BCZ) and the partial ionisation region of heavy elements.

In Section \ref{sec2}, we describe the existing relationship between ionisation and rotation for F-stars. Section \ref{sec3} provides a detailed description of the outer convective envelopes of these stars with an emphasis on the relevant partial ionisation regions. The correlation between the rotational regimes of main-sequence F-stars and the different types of stellar interiors is discussed in Section \ref{sec4}. 
Finally, in Section \ref{sec5} we present a summary and conclusions.

\begin{table*}
	\large
	\begin{threeparttable}
		\caption{Resulting parameters of the theoretical models.}
		\begin{tabular}{ c c c c c c c c c  }          
			\hline
			\hline
			Star Id. &${M}/{M}_\odot$ & ${R}/{R}_\odot$      & ${L}/{L}_\odot$     & $T_{\text{eff}}(K)$      & Age (Gyr)    & $Y_0$  & $\alpha$ & $r_{\text{bcz}}/R$ \\
			\hline
			\hline
			C1 & 0.99    &1.187   & 1.579  & 5943   & 7.259 & 0.275 & 1.70 & 0.740 \\
			C2 & 1.20    &1.271   & 2.404  & 6383   & 3.782 & 0.253 & 2.30 & 0.775 \\
			C3 & 1.24    &1.409   & 2.470  & 6107   & 3.802 & 0.248 & 1.65 & 0.803 \\ 
			C4 & 1.05    &1.210   & 1.864  & 6136   & 6.459 & 0.260 & 2.00 & 0.732 \\ 
			C5 & 1.20    &1.424   & 3.036  & 6392   & 2.982 & 0.287 & 1.80 & 0.847 \\
			\hline
			H1 & 1.55    & 2.195 & 6.984  & 6338  & 2.182 & 0.249 & 1.50 & 0.929  \\
			H2 & 1.45    & 1.960 & 5.979  & 6453  & 1.982 & 0.313 & 1.90 & 0.871  \\
			H3 & 1.39    & 1.665 & 4.146  & 6388  & 2.582 & 0.265 & 1.90 & 0.839  \\
			H4 & 1.50    & 1.572 & 6.741  & 6741  & 0.472 & 0.326 & 1.80 & 0.951  \\
			H5 & 1.39    & 2.033 & 6.546  & 6481  & 2.782 & 0.278 & 1.75 & 0.881  \\
			\hline
			\hline
		\end{tabular}
		\label{table:2}
	\end{threeparttable}
\end{table*}

\section{Ionization and rotation: a relation unveiled by the seismology of the outer layers}\label{sec2}

\begin{figure*} 
	\includegraphics[width=1.0\textwidth]{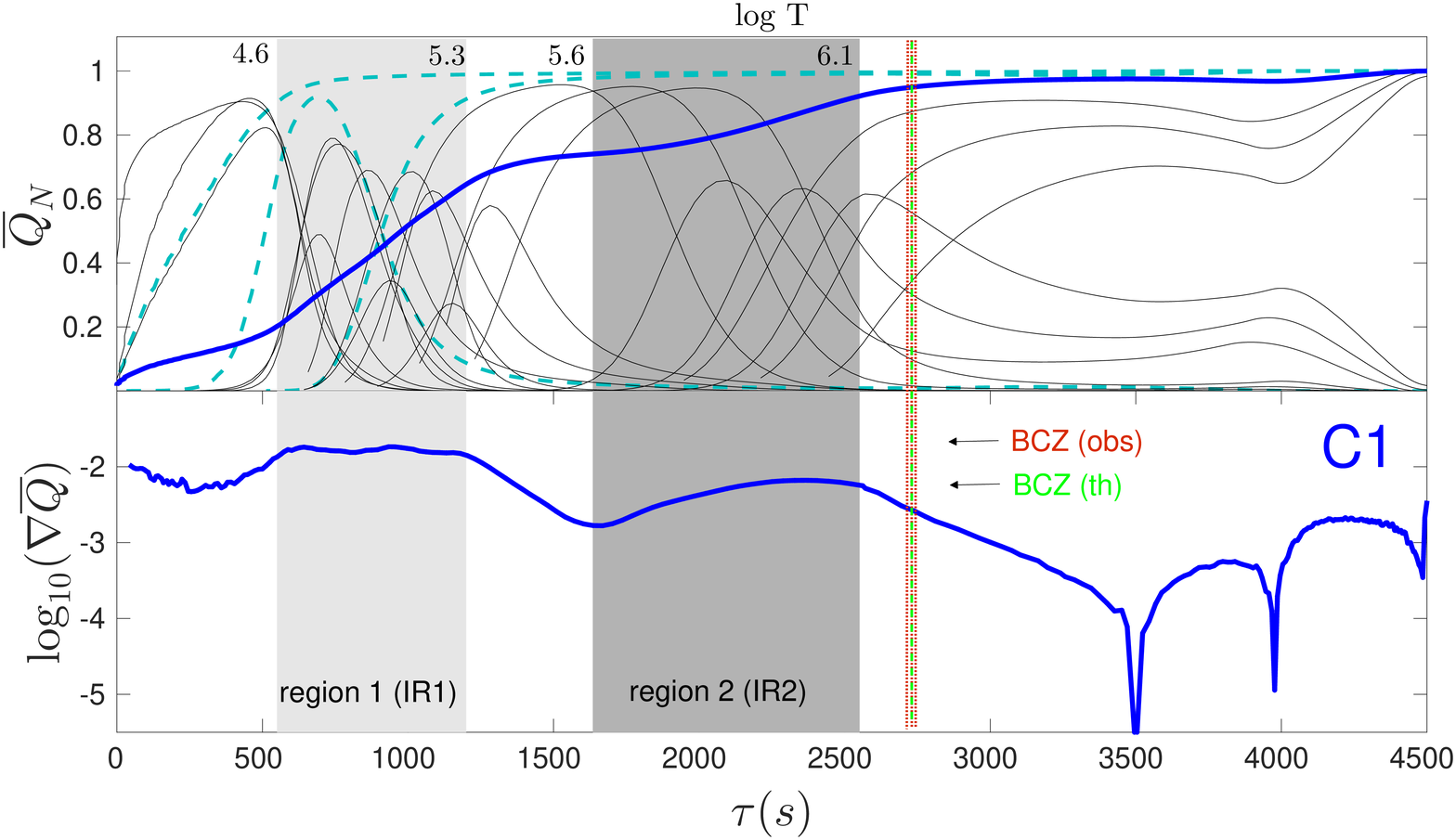}\\
	\includegraphics[width=1.0\textwidth]{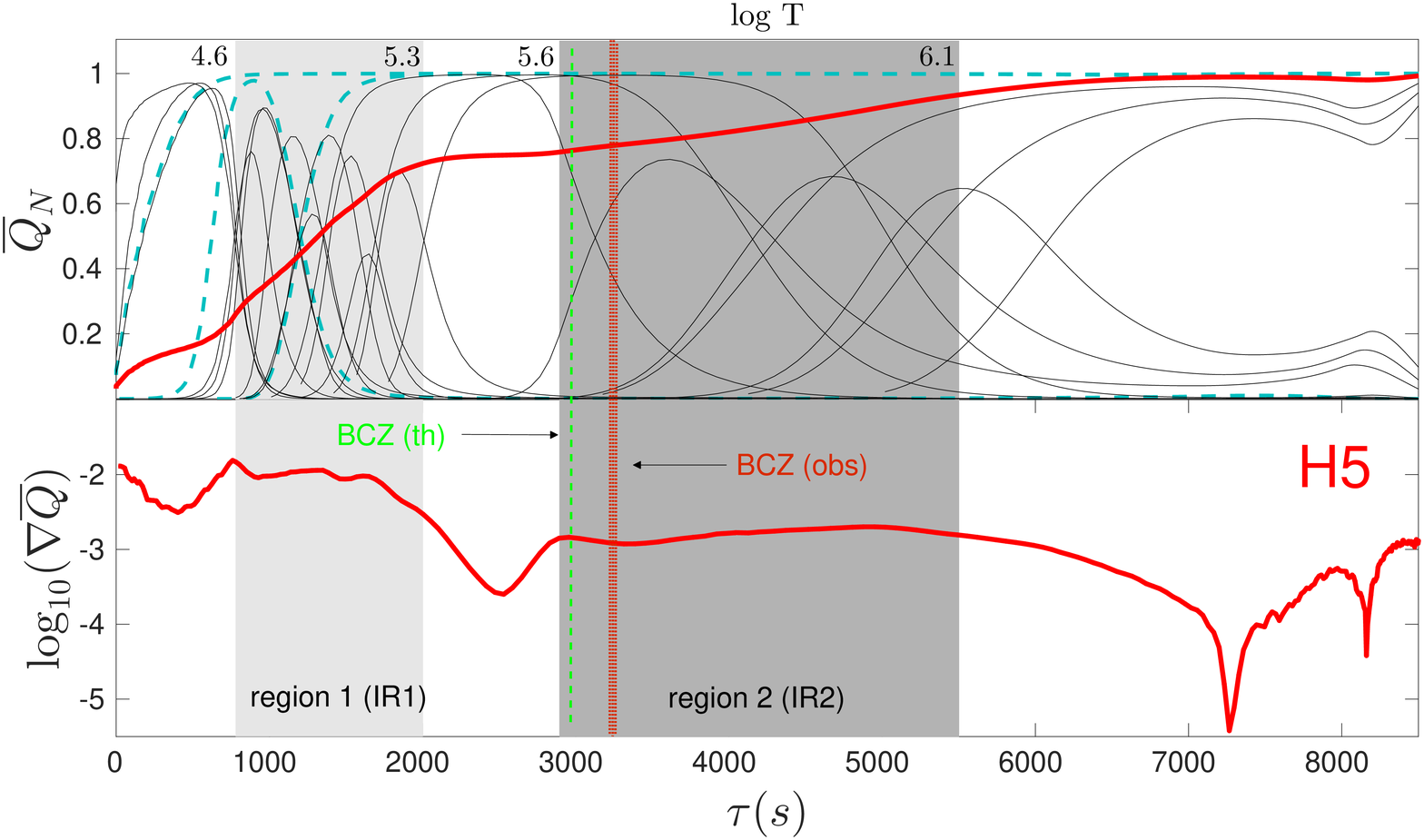}\\
	\caption{Detailed representation of the partial ionisation processes occurring in the interior of a cool star (top; blue colour) and in the interior of a hot star (bottom; red colour). The ionisation fractions for light elements (H, He) are indicated with dashed lines, whereas the ionisation fractions for heavy elements are indicated with thin solid black lines \citep[see][]{2017MNRAS.466.2123B}. The mean effective ionic charge normalised to its maximum value, $\overline{Q}_N = \overline{Q}/\max{\overline{Q}}$, is given by the thick solid lines (blue, red) in each case. Below, the corresponding gradient of the ionic charge is also given by a thick solid line in each case. The partial ionisation regions IR1 and IR2 represent the locations where the effective mean ionic charge increases significantly. IR1 coincides with the partial ionisation region of helium and IR2 is associated with partial ionisation of heavier elements. Both these regions are marked with vertical grey bars. Also shown are the locations of the base of the convective envelopes for both stars. The red vertical dashed line represents the observed value. The thickness of the line indicates $\tau_{\text{bcz}}$ approximately within 1-$\sigma$ error bar. Dashed vertical green line stands for the theoretical value of $\tau_{\text{bcz}}$.}
	\label{fig4}
\end{figure*}

\begin{figure*} 
	\includegraphics[width=1.0\textwidth]{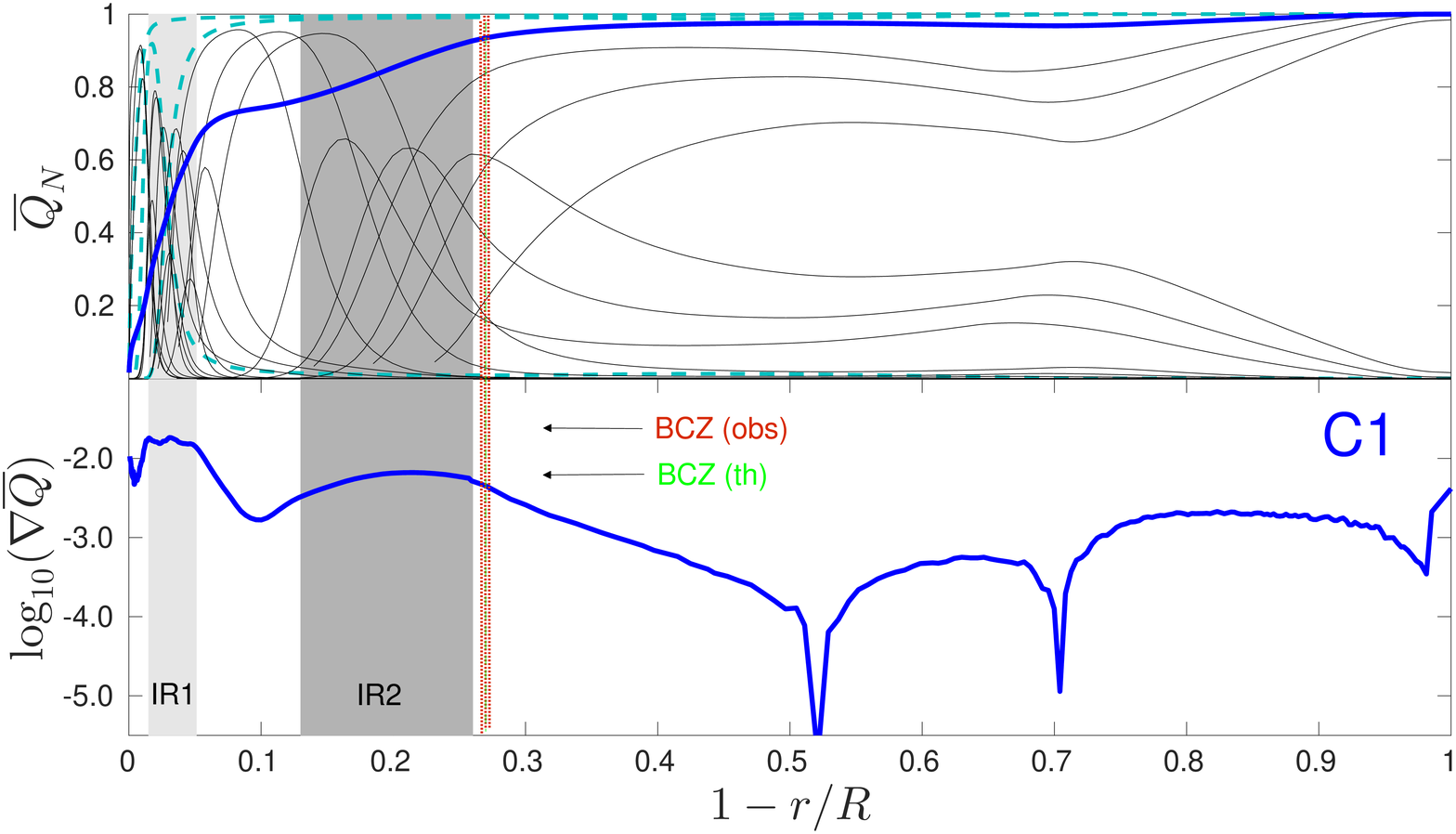}\\
	\includegraphics[width=1.0\textwidth]{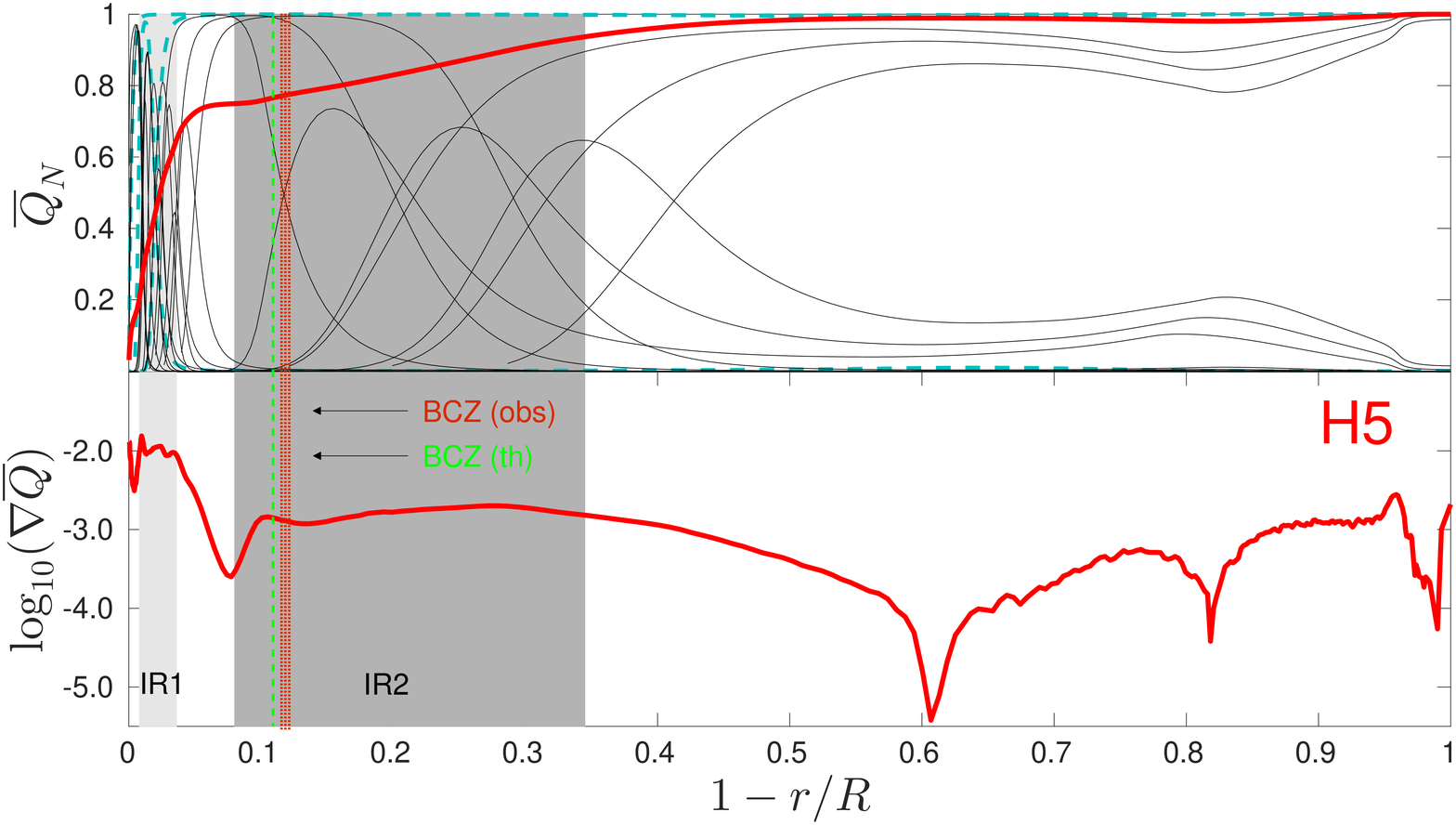}\\
	\caption{Detailed representation of the partial ionisation processes occurring in the interior of a cool star (top; blue colour) and in the interior of a hot star (bottom; red colour). The ionisation fractions for light elements (H, He) are indicated with dashed lines, whereas the ionisation fractions for heavy elements are indicated with thin solid black lines \citep[see][]{2017MNRAS.466.2123B}. The mean effective ionic charge normalised to its maximum value, $\overline{Q}_N = \overline{Q}/\max{\overline{Q}}$, is given by the thick solid lines (blue, red) in each case. Below, the corresponding gradient of the ionic charge is also given by a thick solid line in each case. The partial ionisation regions IR1 and IR2 represent the locations where the effective mean ionic charge increases significantly. IR1 coincides with the partial ionisation region of helium and IR2 is associated with partial ionisation of heavier elements. Both these regions are marked with vertical grey bars. Also shown are the locations of the base of the convective envelopes for both stars. The red vertical dashed line represents the observed value. The thickness of the line indicates $\tau_{\text{bcz}}$ approximately within 1-$\sigma$ error bar. Dashed vertical green line stands for the theoretical value of $\tau_{\text{bcz}}$. All quantities are plotted as a function of the star's fractional radius.}
	\label{fig5}
\end{figure*}

\begin{figure*} 
	\includegraphics[width=0.49\textwidth]{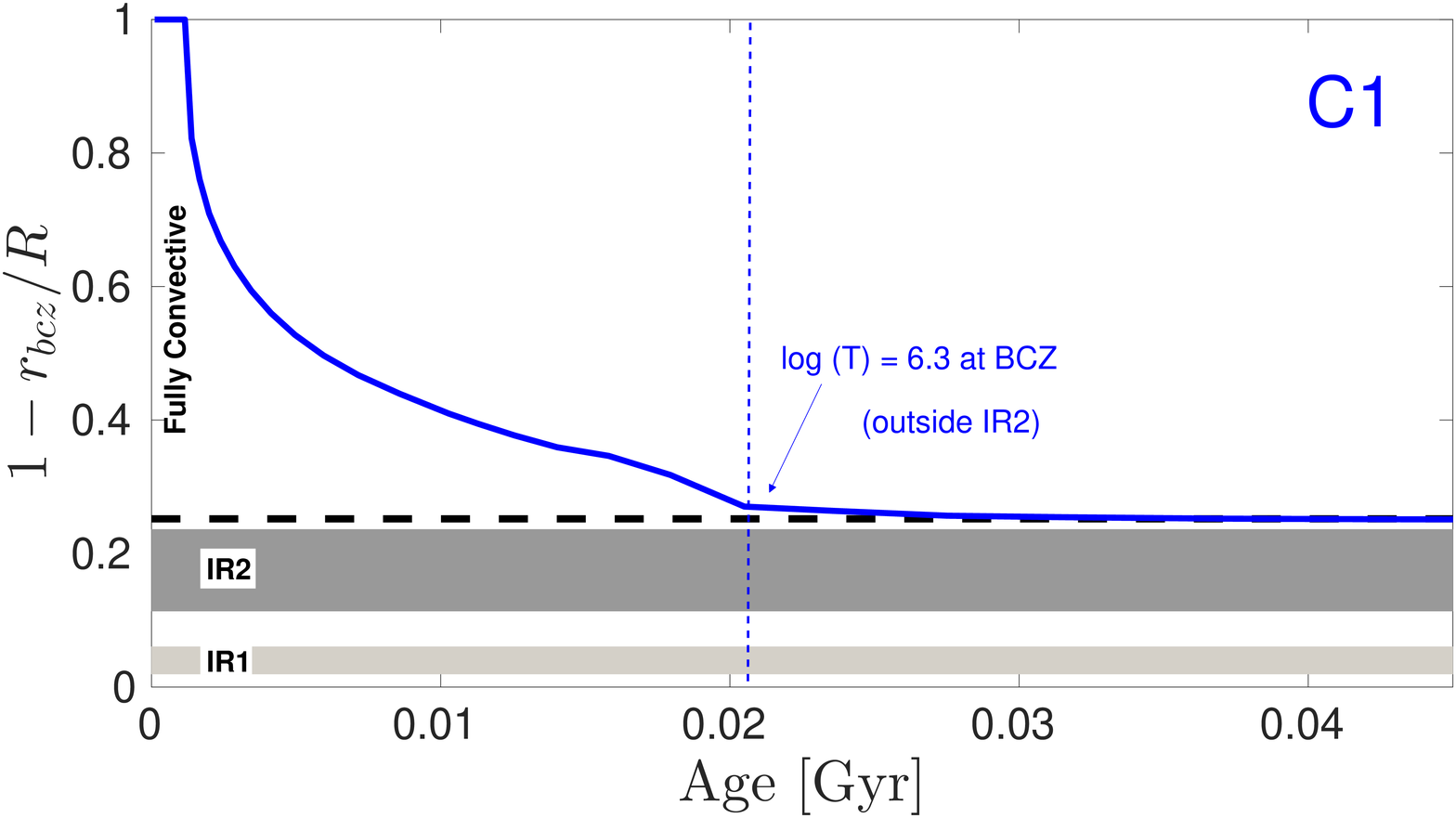}
	\includegraphics[width=0.49\textwidth]{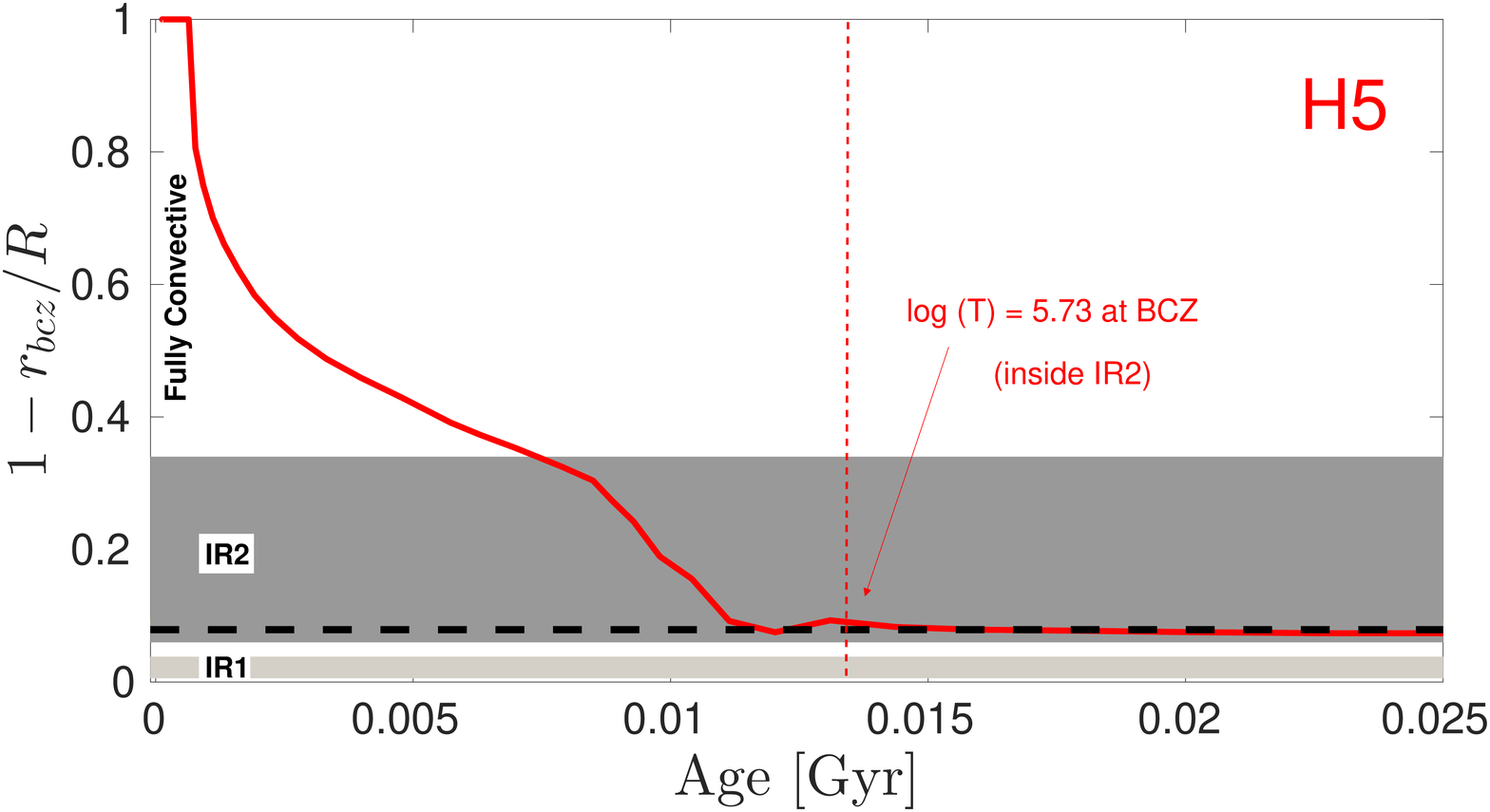}
	\caption{Solid lines, blue and red, represent the evolution of the location of the base of the convective zone from the fully-convective pre-main-sequence phase up to a few gigayears after the ZAMS (zero-age main sequence). From ZAMS to the final age of the model, the location of the BCZ keeps almost unchanged. The horizontal dashed line indicates the approximate location for  the BCZ during all the main-sequence evolution time. Particularly important are the dashed vertical lines, red and blue, which represent the moment of the zero age main sequence. Indicated is also the value of the temperature at the base of the convective zone for the corresponding model age.}
	\label{fig6}
\end{figure*}

\begin{figure*}
	\centering
	\includegraphics[width=0.7\textwidth]{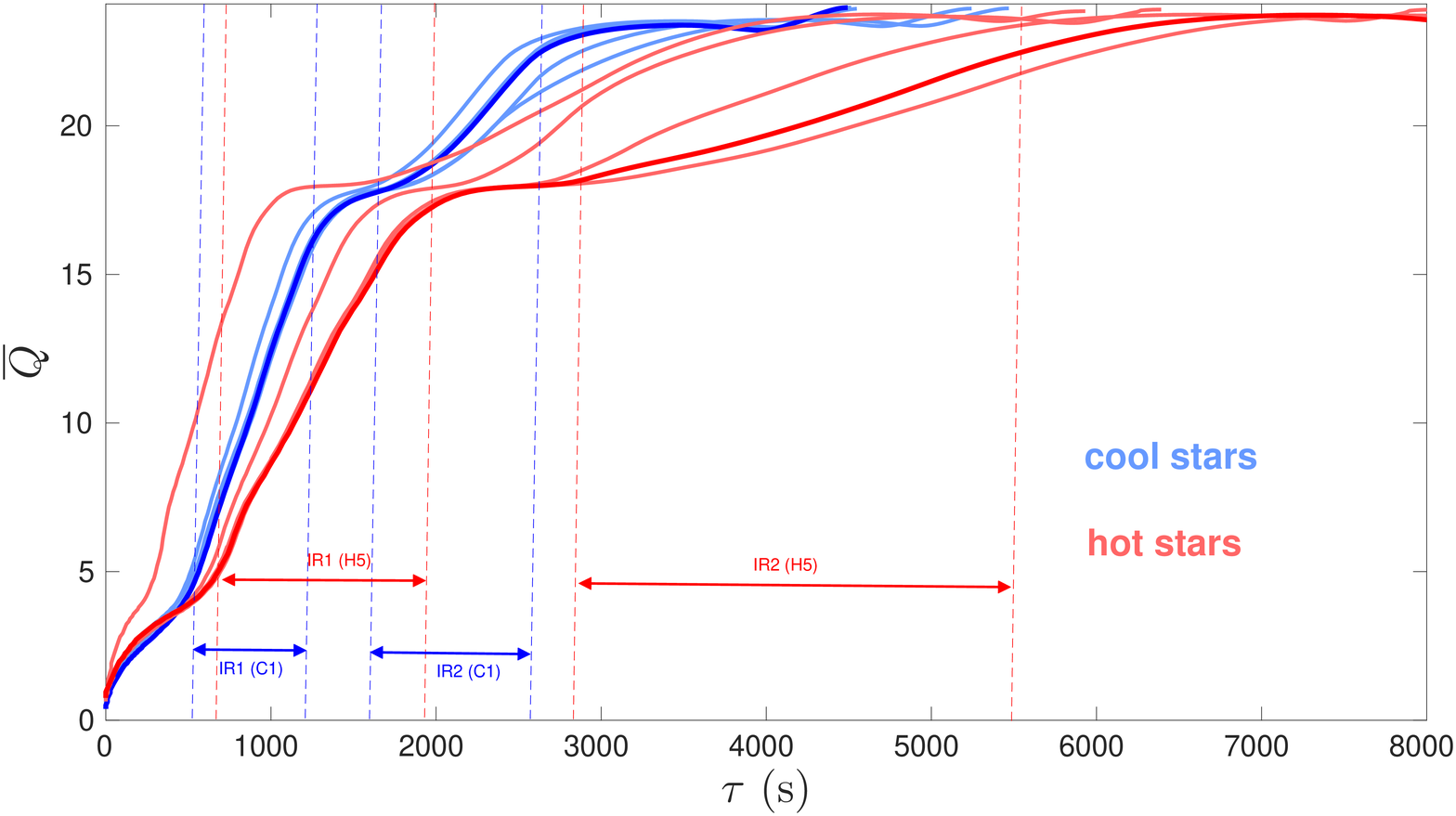}
	\caption{The mean effective ionic charge as a function of the acoustic depth for all the theoretical models.  Blue colour for cool stars and red colour for hot stars. Highlighted are the ionisations regions IR1 and IR2 for the models C1 and H5.}
	\label{fig7}
\end{figure*}

F-stars are expected to have unstable outer regions where the energy transport is dominated by convection; hot portions of plasma ascend to the surface whereas cold portions sink towards the centre. The depths of these convective envelopes is diverse and depends on the mass and chemical abundances \citep[e.g.,][]{2012ApJ...746...16V}. Hot stars, being more massive, have usually thiner convective zones than cool stars.
The convective external layers of these stars are regions of great uncertainty because in these regions coexist a multitude of physical processes taking place at the same timescales, such as for instance, the turbulent motions of convection itself, magnetic fields, partial ionisation processes and the excitation and damping of solar-like oscillations. Therefore, the outer layers are complex and difficult to study.

\begin{figure*}
	\includegraphics[width=0.49\textwidth]{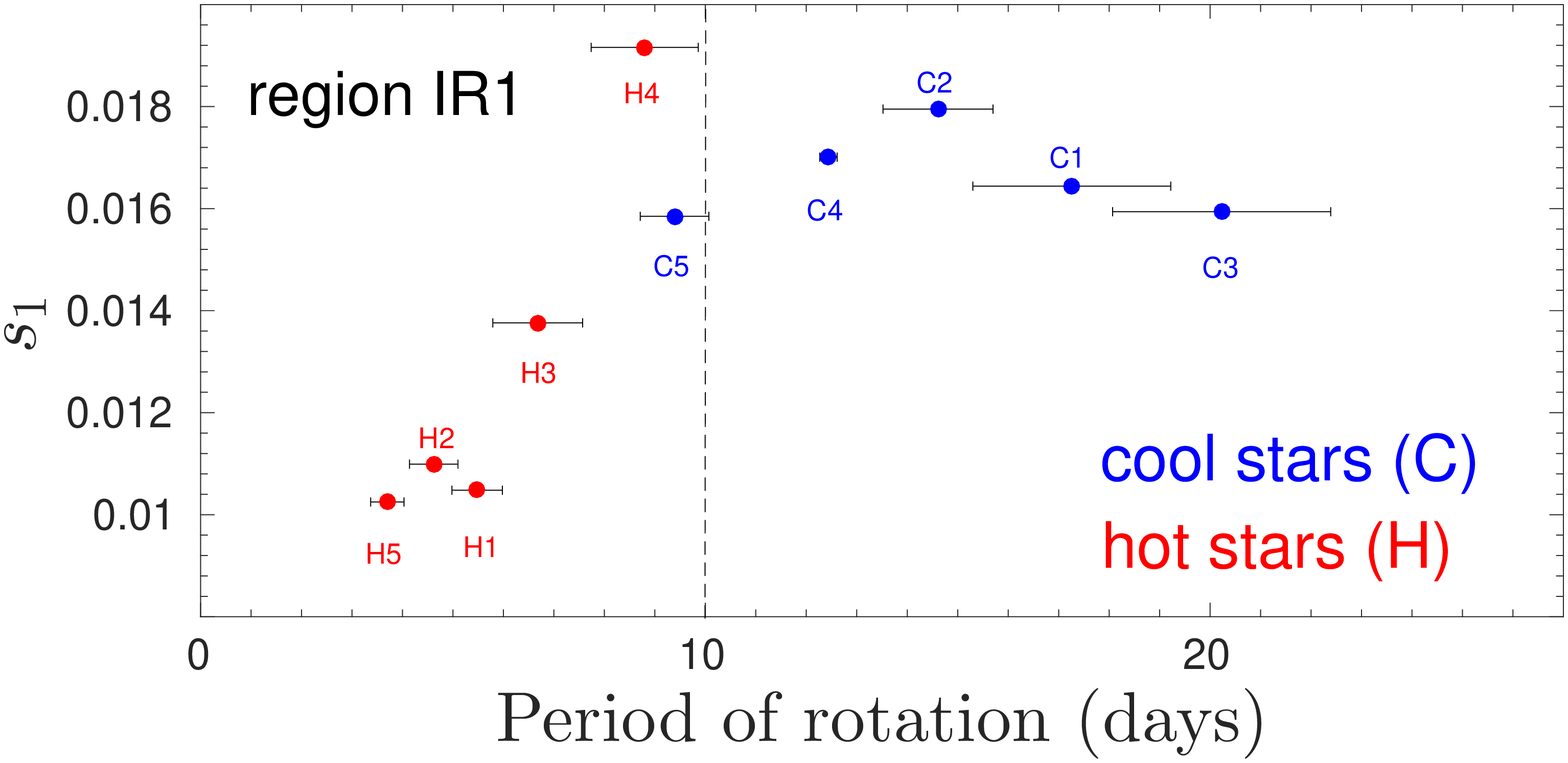}
	\includegraphics[width=0.49\textwidth]{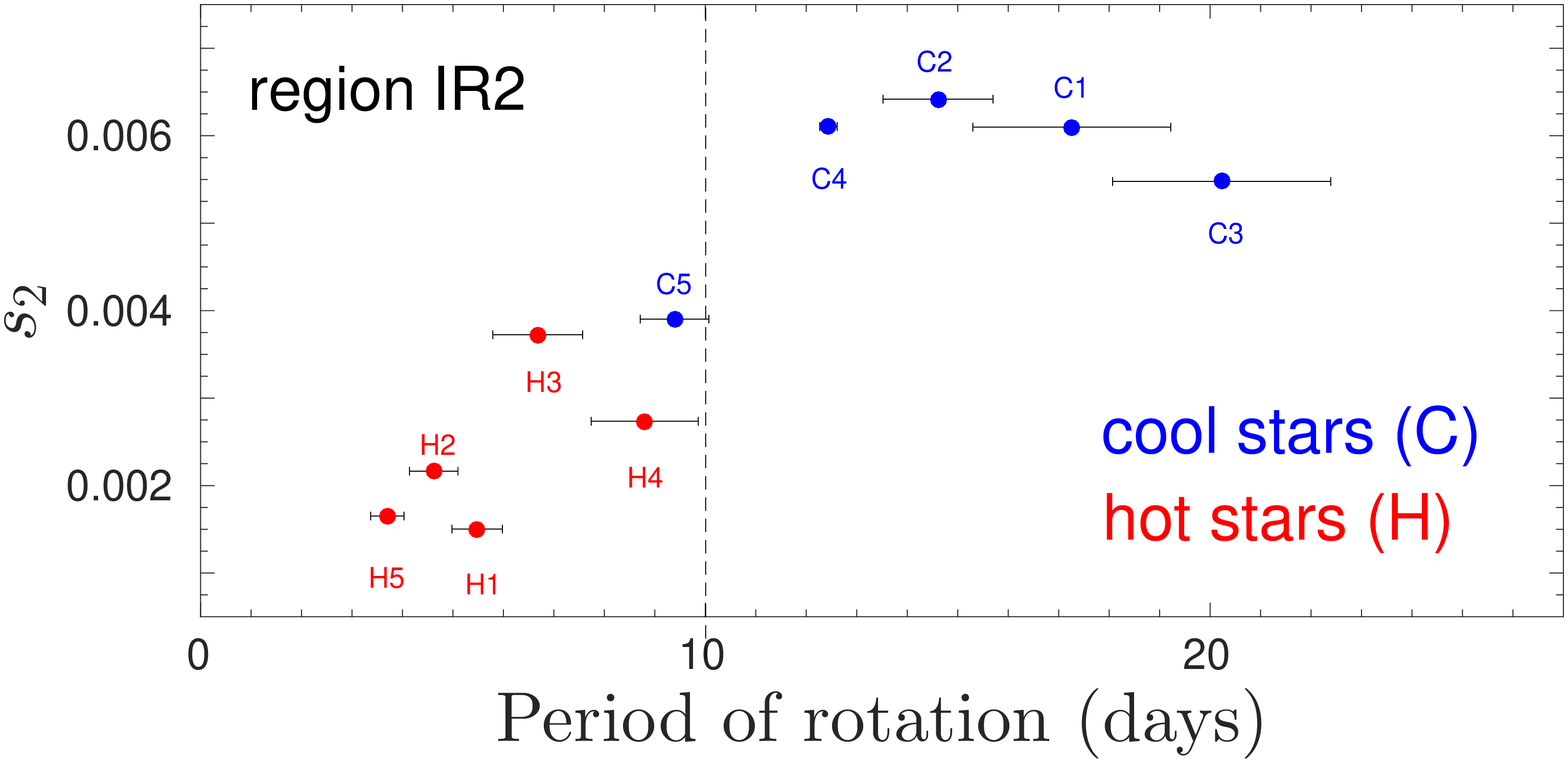}\\
	\includegraphics[width=0.49\textwidth]{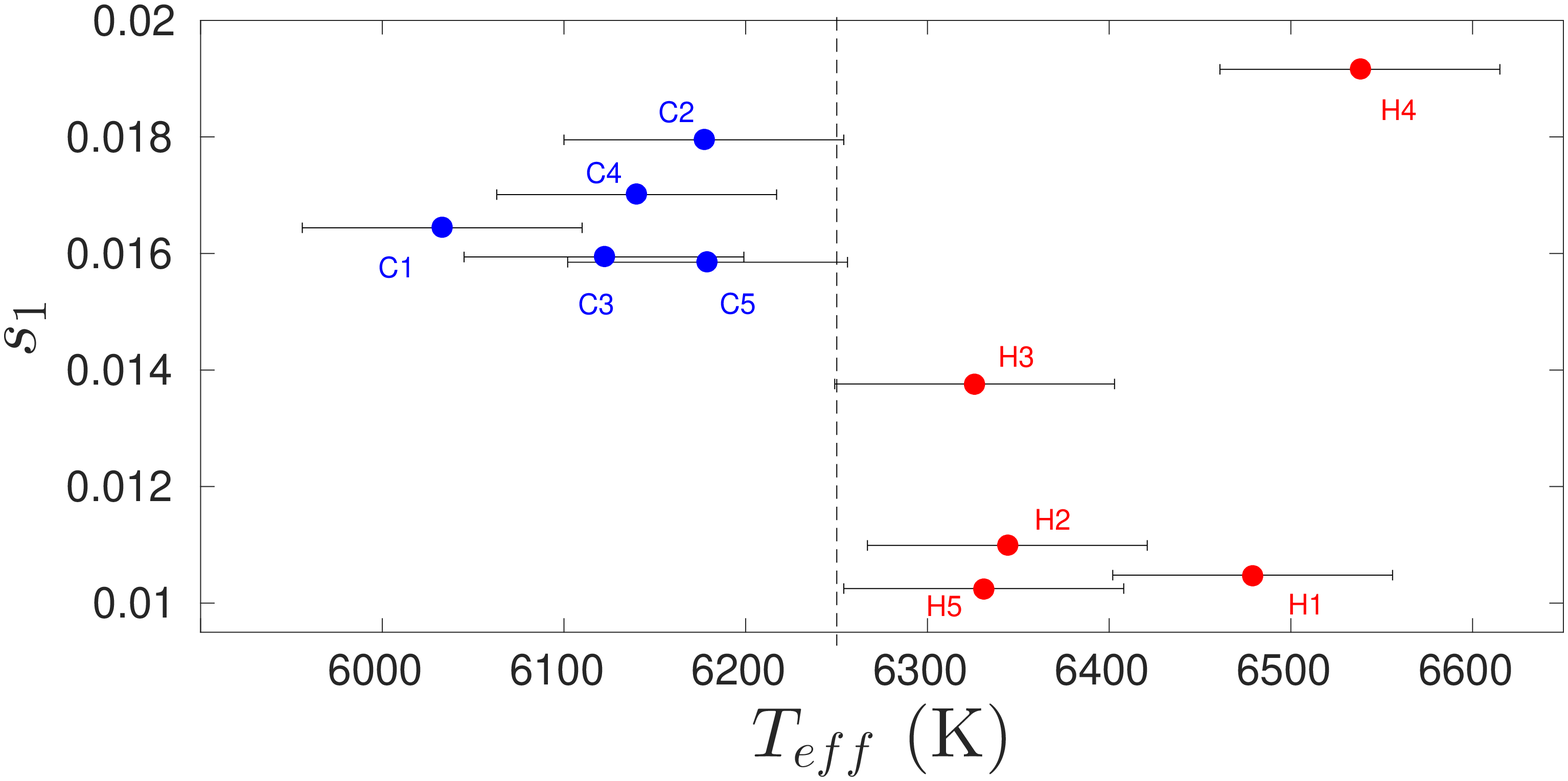}
	\includegraphics[width=0.49\textwidth]{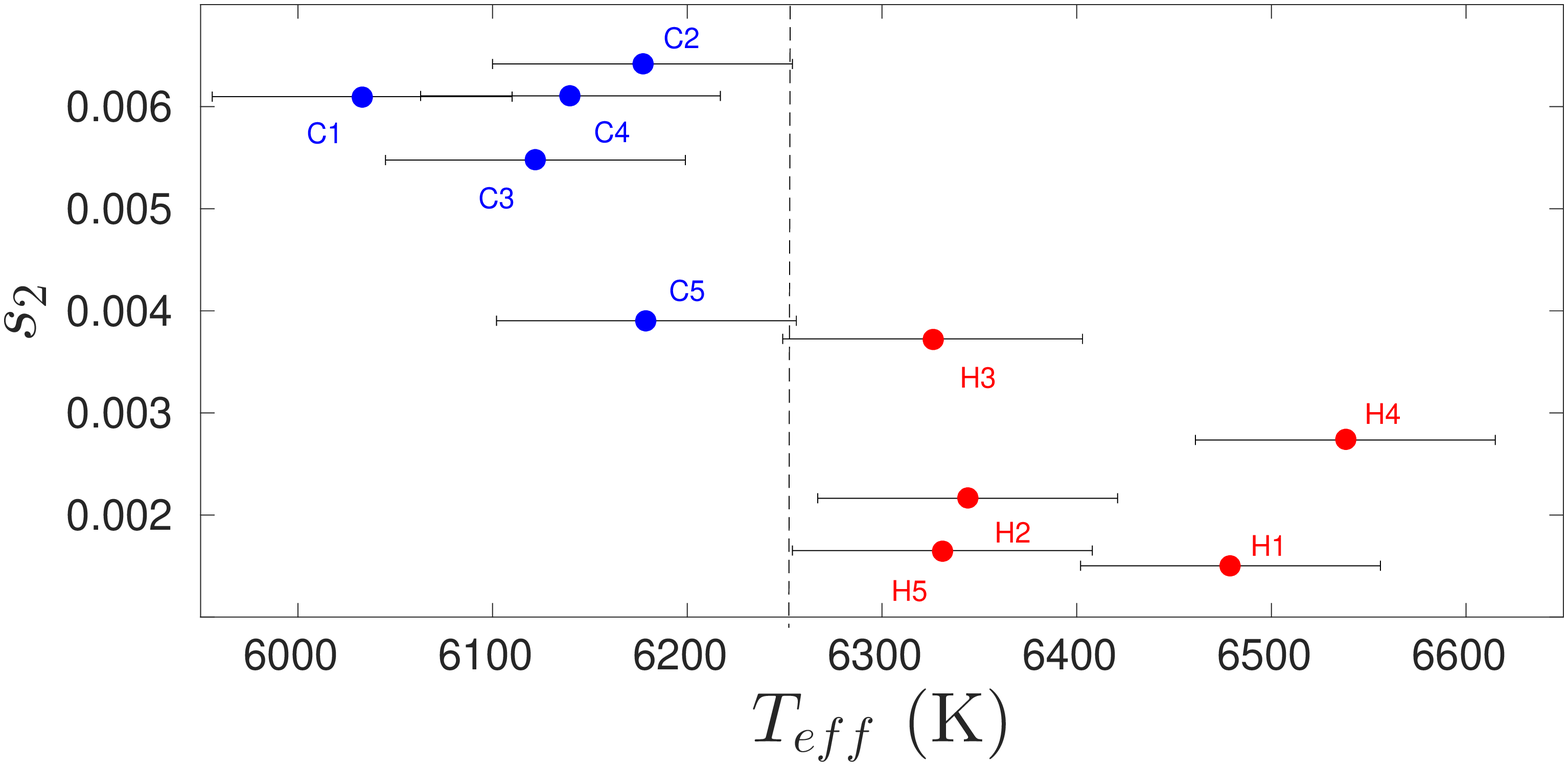}\\
	\includegraphics[width=0.49\textwidth]{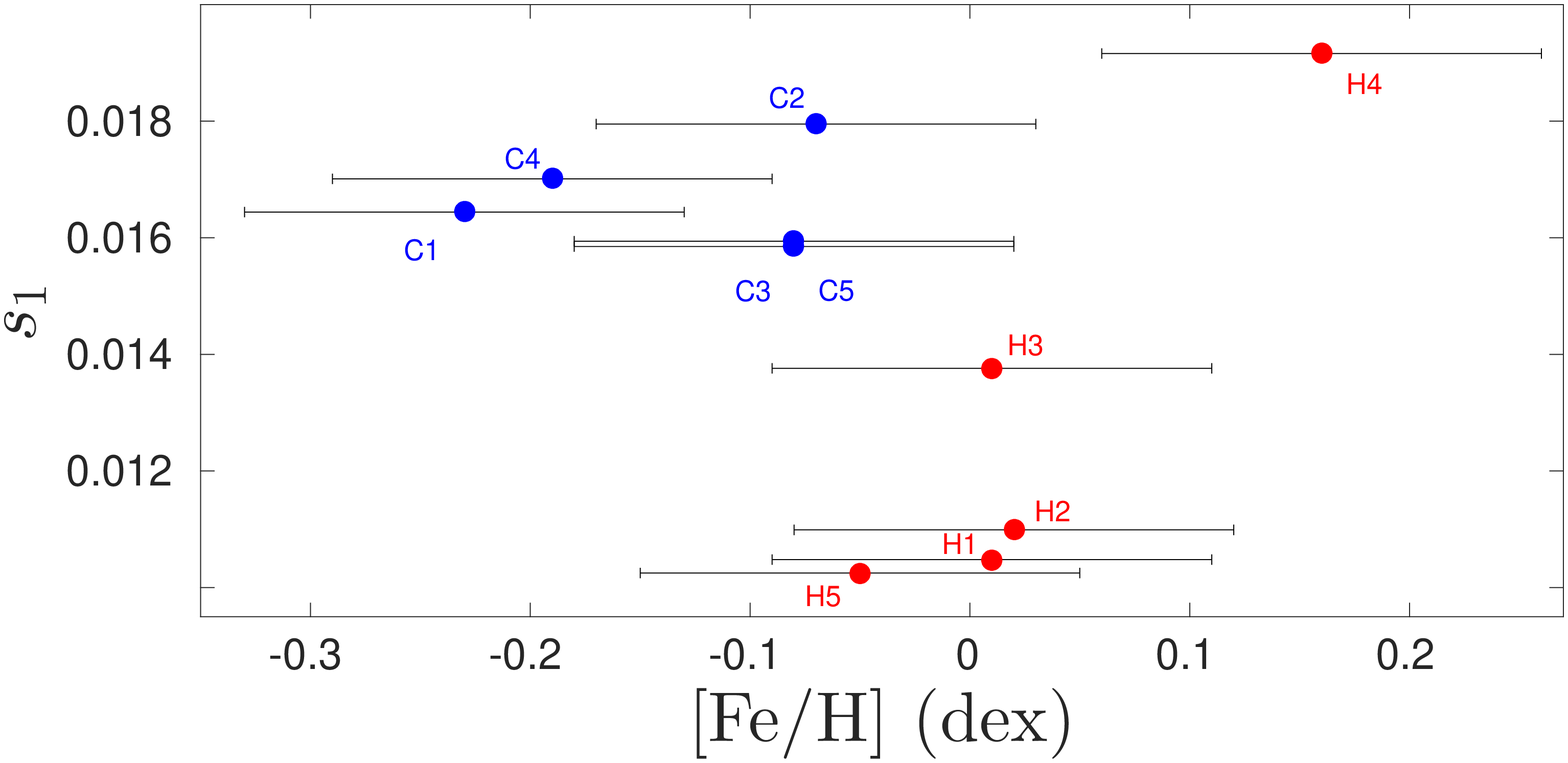}
	\includegraphics[width=0.49\textwidth]{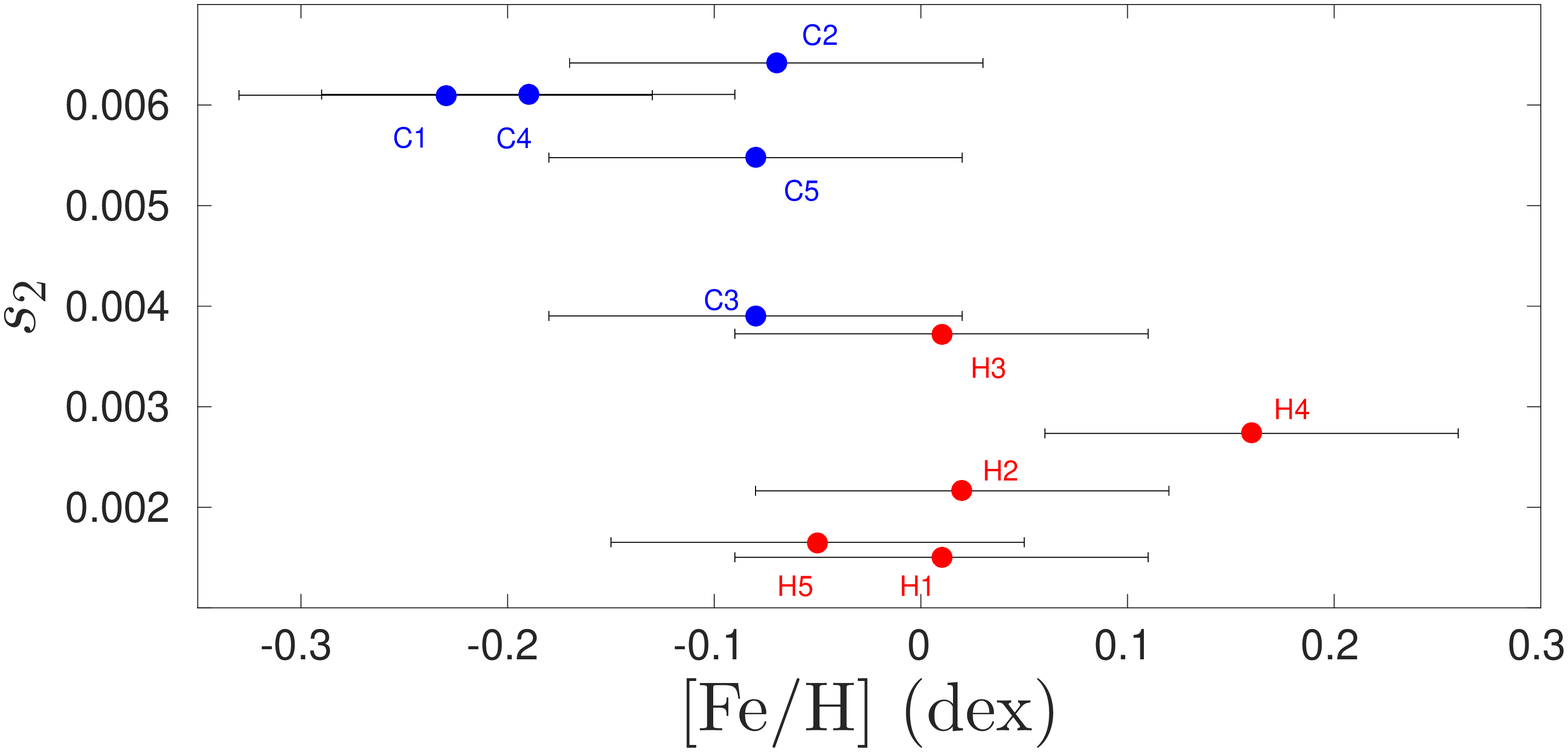}\\
	\includegraphics[width=0.49\textwidth]{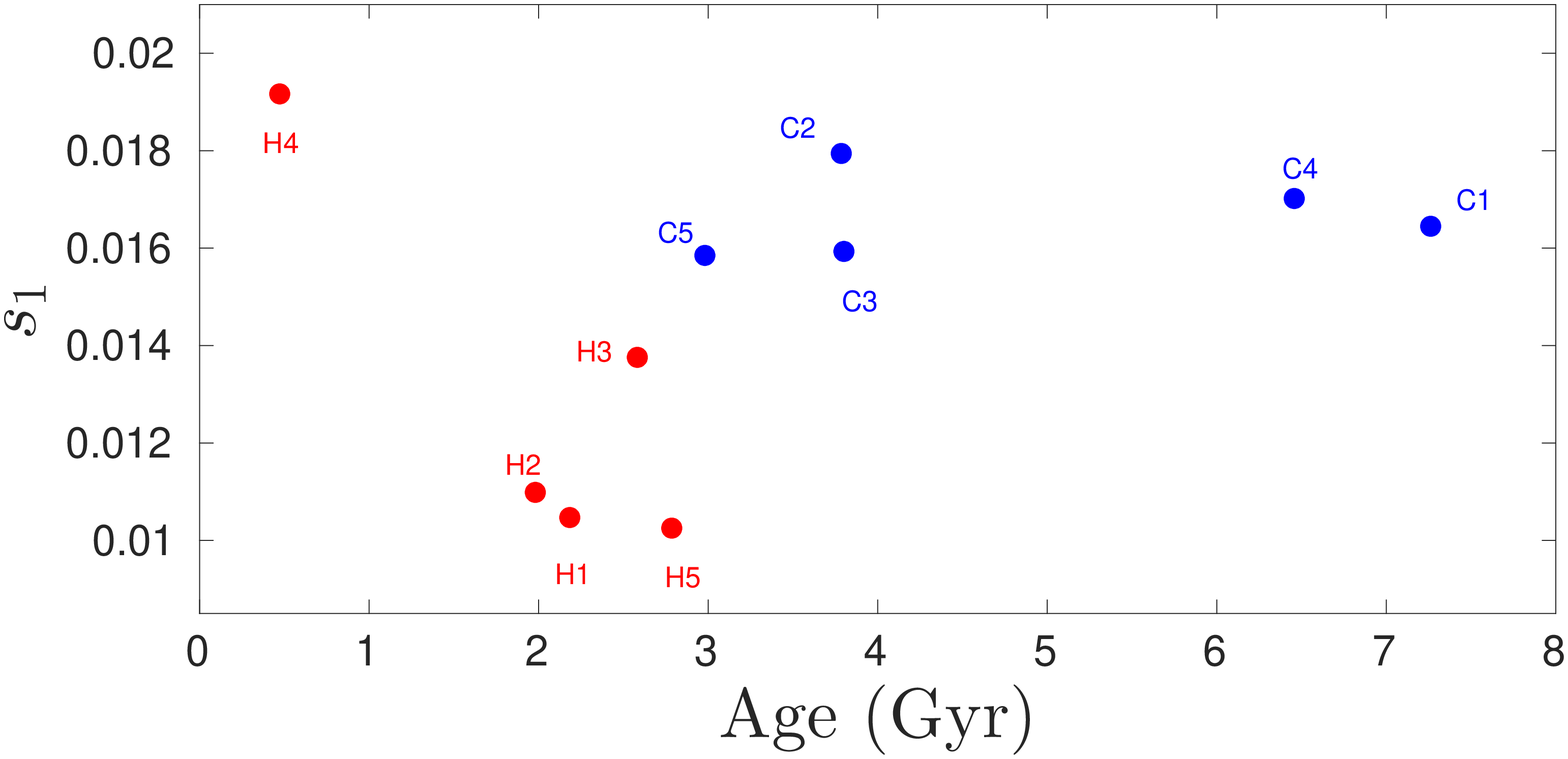}
	\includegraphics[width=0.49\textwidth]{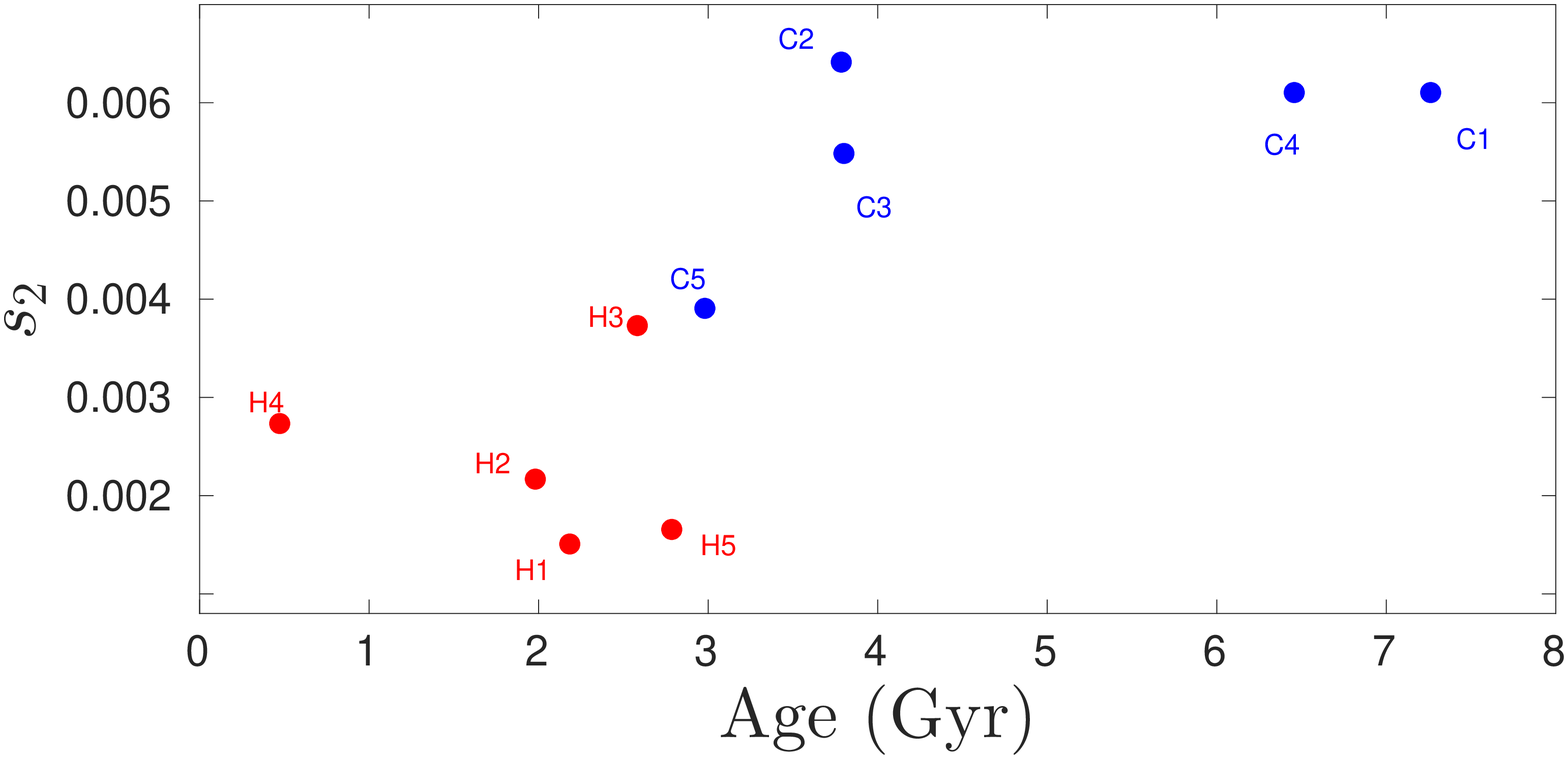}\\
	\includegraphics[width=0.49\textwidth]{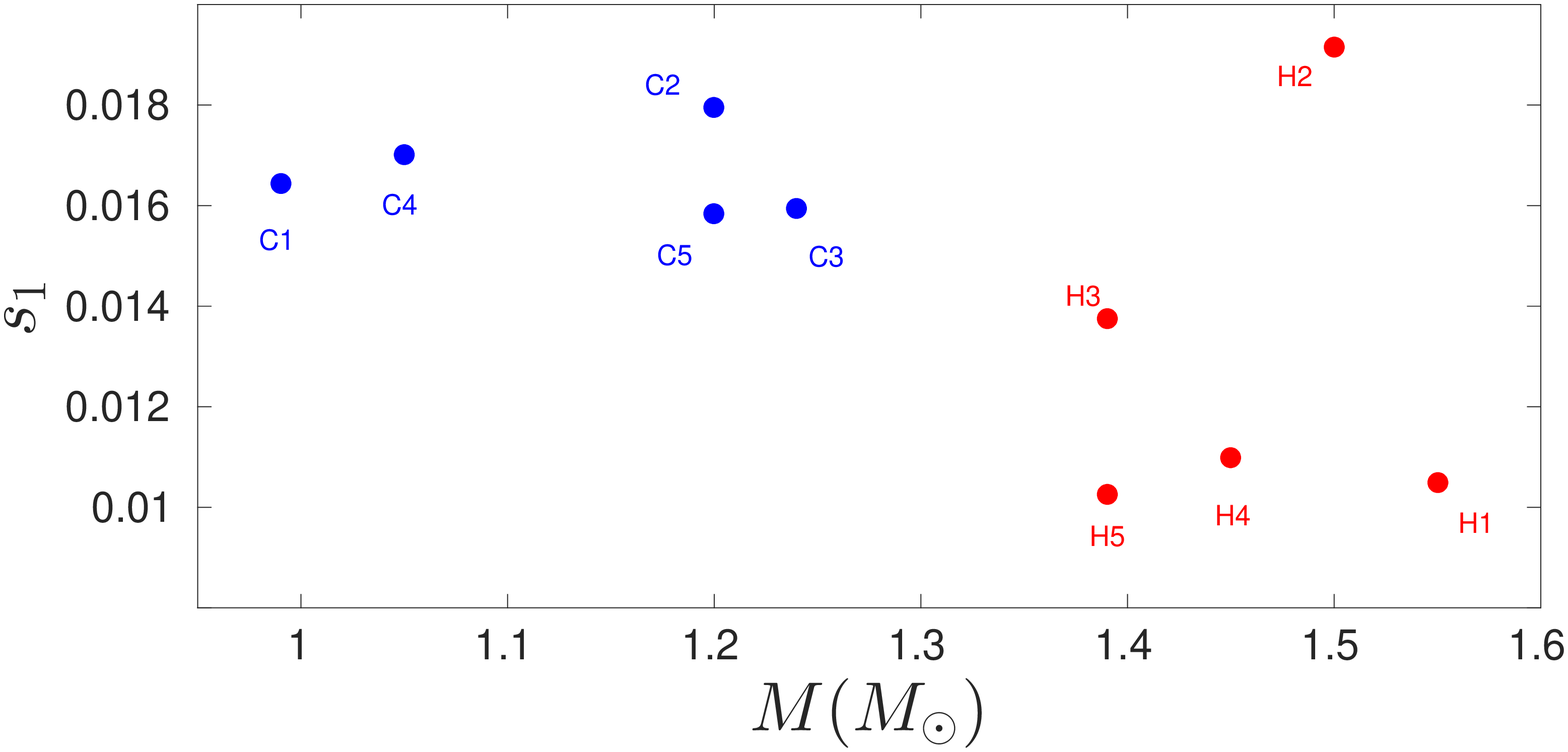}
	\includegraphics[width=0.49\textwidth]{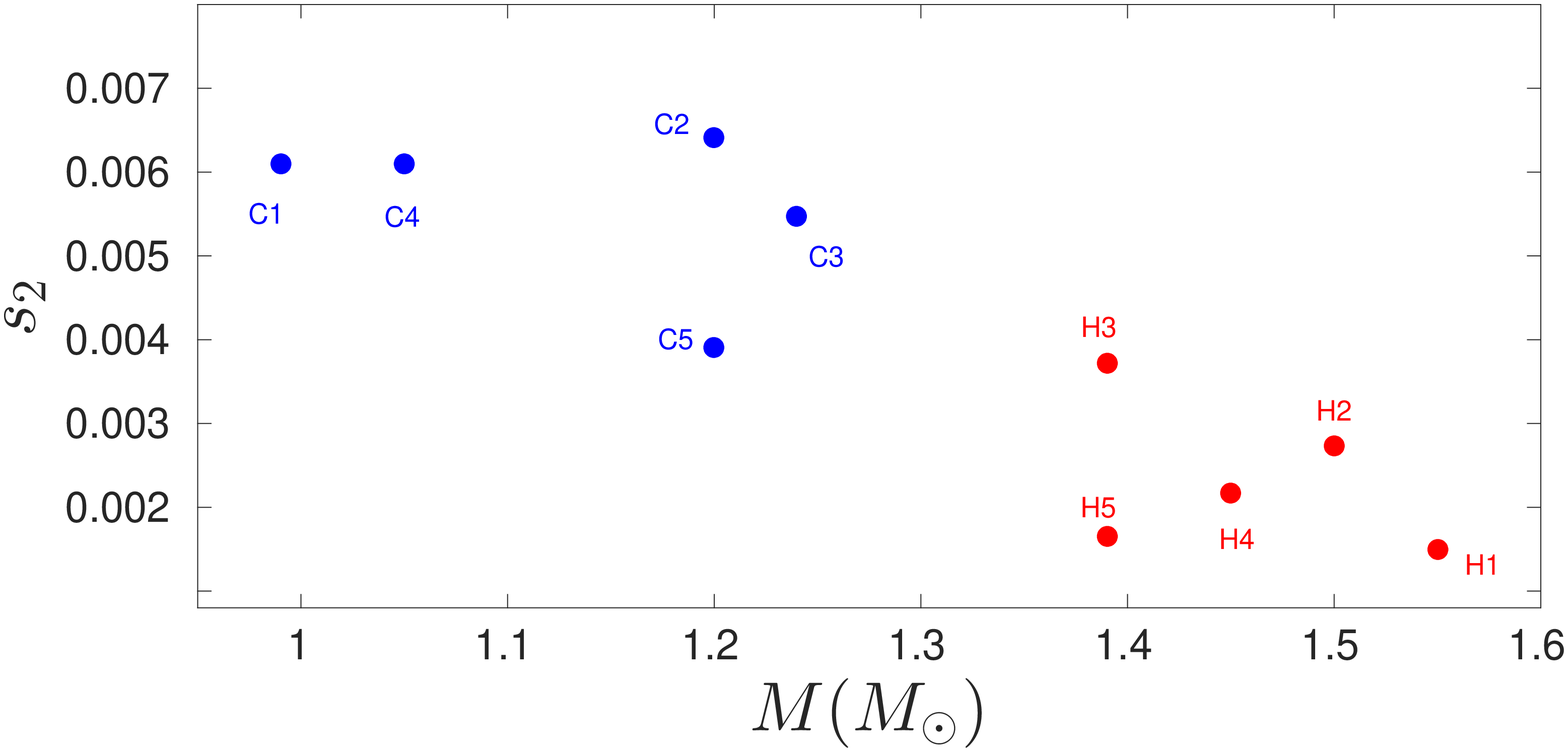}
	\caption{Left panel: the slopes associated with the increase of the mean ionic charge due to ionisation processes in region IR1 ($s_1$) plotted against fundamental stellar properties; rotation period, effective temperature, metallicity, age, and mass. Right panel: the slopes associated with the increase of the mean ionic charge due to ionisation processes in region IR2 ($s_2$) plotted against fundamental stellar properties; rotation period, effective temperature, metallicity, age, and mass. Blue colour for cool stars and red colour for hot stars. Vertical dashed lines indicate the value of the rotation period and effective temperature around which occurs the transition in the main sequence rotational regimes--the Kraft break.}
	\label{fig8}
\end{figure*}

\begin{figure*} 
	\includegraphics[width=0.7\textwidth]{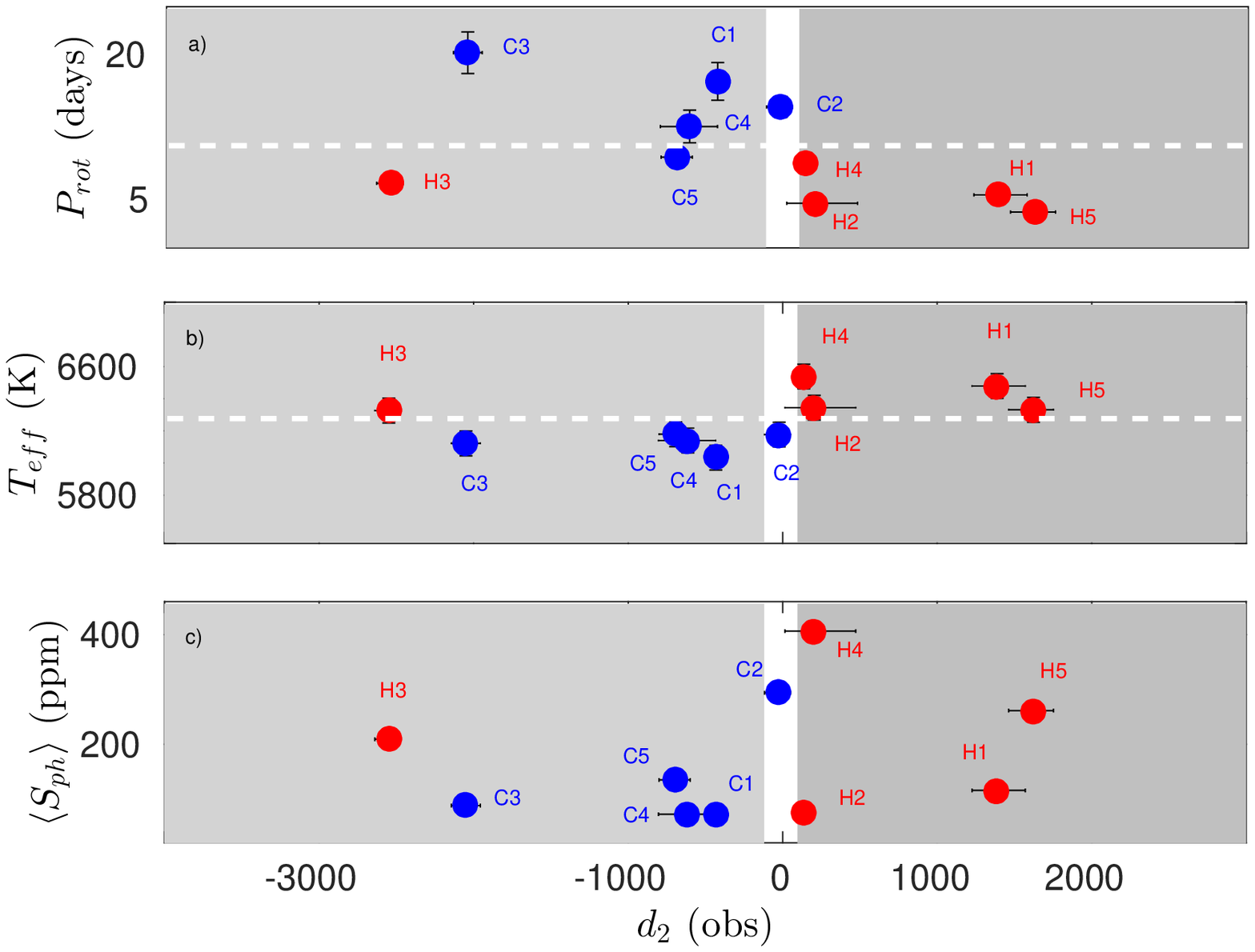}\\
	\includegraphics[width=0.7\textwidth]{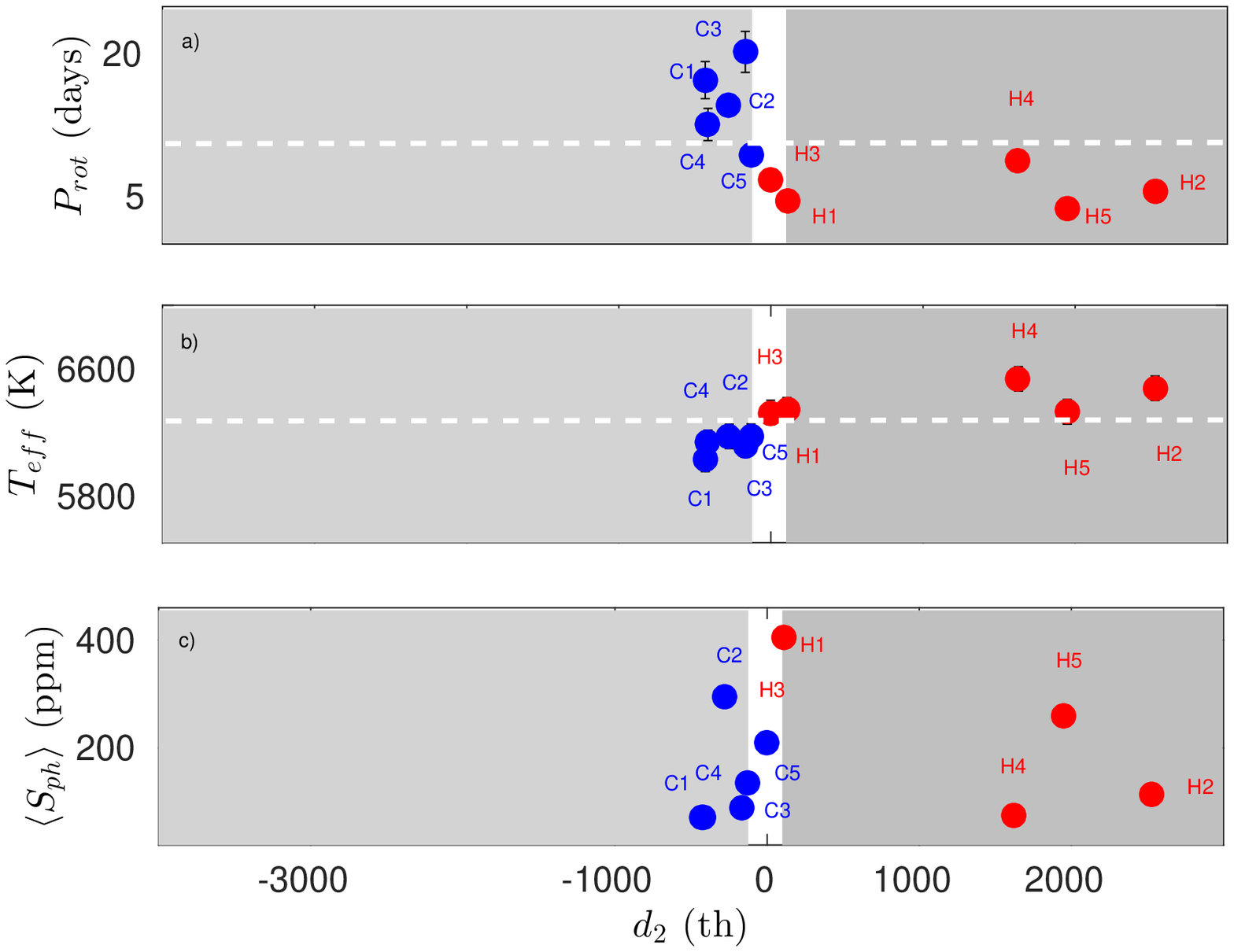}
	\caption{$d_2$ represents the difference between the acoustic depths of the location of the region IR2 and the acoustic depth of the location of the BCZ (Eq. \ref{eq4}).  A negative value means that the location of the region IR2 is above the base of the convective zone, i.e., IR2 is closer to the surface than the BCZ. A zero value happens when the two zones coincide. Finally, a positive value means the region IR2 is located (mostly) below the BCZ.  The white vertical bar marks the overlap of IR2 (central point) and BCZ. White horizontal dashed lines correspond to the threshold values of the Kraft break for the rotation period ($\sim 10$ days) and $T_{\text{eff}}$ ($\sim 6250$ K). $d_2$ is plotted against: a) the rotation period ($P_{\text{rot}}$); b) the effective temperature ($T_{\text{eff}}$) and c) the magnetic photometric index ($S_{\text{ph}}$). Top panel: was used the observational value for the location of the base of the convective zone, $\tau_{\text{BCZ}}$ \citep{2017ApJ...837...47V}. Bottom panel: was used the theoretical value for the location of the base of the convective zone, $\tau_{\text{BCZ}}$.}
	\label{fig9}
\end{figure*}

Asteroseismology, as it is well known, offers the possibility to probe the internal structure of a star by using stellar pulsation data.  Particularly, regarding the outer layers, the signatures of the acoustic sharp features, i.e., acoustic glitches, can be used to find the locations of the base of the convective zones and also of the ionisation zones of helium \citep[e.g.,][]{1990LNP...367..283G, 2001A&A...368L...8M, 2004A&A...423.1051B, 2007MNRAS.375..861H, 2014ApJ...782...18M, 2017ApJ...837...47V}.

The well-known Duvall law \citep{1982Natur.300..242D}
\begin{equation}\label{eq1}
F(W) = \pi \left(\frac{\alpha(\omega) + n}{\omega}   \right) \, ,
\end{equation}
where $W=\omega / L$, $L=l+1/2$, and $n$ and $l$ are, respectively, the radial order and the degree of the acoustic mode, provides a bridge between theoretical and observational data. Here, $\omega=2\pi\nu$, represents the angular frequency of the acoustic mode, whereas $\nu$ stands for the cyclic frequency. The left-hand side of equation \ref{eq1} is determined by the radial distribution of the sound speed, i.e., contains information about the structural model parameters. The right-hand side is, in turn, of observational nature, with $\alpha(\omega)$ being the dependence on the frequency of the phase shift of the reflected acoustic waves. This phase shift absorbs the difference between the first-order asymptotic eigenfrequency equation \ref{eq1} and the full equation given by the asymptotic theory of adiabatic non-radial oscillations \citep[e.g.,][]{1979nos..book.....U}. After some manipulation of equation \ref{eq1} \citep[e.g.,][]{1987SvAL...13..179B} is possible to obtain two relations involving the phase shift $\alpha(\omega)$:
\begin{equation}\label{eq2}
\beta(\omega) = - \omega^2 \frac{d}{d \omega} \left( \frac{\alpha(\omega)}{\omega} \right)
\end{equation}
and
\begin{equation}\label{eq3}
\beta(\omega) = \frac{\omega - n \left( \frac{\partial \omega}{\partial n}\right) - L\left( \frac{\partial \omega}{\partial L}\right) }{\left( \frac{\partial \omega}{\partial n}\right)} \, .
\end{equation}
These two relations define a robust seismic diagnostic that can be determined from theoretical data and also from observational data.
The seismic parameter $\beta(\omega)$ acts, then, as a proxy for the phase shift of the reflected by the surface acoustic waves \citep[e.g.,][]{1989ASPRv...7....1V, 1994A&A...290..845L, 2001MNRAS.322..473L, 2014ApJ...782...16B}.  
This parameter is specifically suited to study the outer convective layers of main-sequence stars and remarkably sensitive to partial ionisation processes taking place in these layers. 
Nevertheless, we note that the observational frequencies are obtained from low-degree modes and thus influenced by the effect of the gravitational potential. Theoretical data, in turn, is obtained in the Cowling approximation which neglects the effect of the gravitational potential. The result is a well-know systematic l-dependence in the $\beta(\omega)$ diagnostic \citep[e.g.,][]{1997ApJ...480..794L}.

In the Sun's case this diagnostic afforded the measurement of the solar helium abundance \citep{1991Natur.349...49V} and was able to contribute to the calibration of the equation of state \citep{1991A&A...248..263P, 1992MNRAS.257...32V}. Concerning other stars it begins now to reveal its potential thanks to the quantity and quality of the individual frequency modes made available by the space-based missions. Figure \ref{fig1} shows the seismic parameter $\beta(\nu)$ (where $\nu=\omega/2\pi$) computed from tables of observational frequencies \citep{2012A&A...543A..54A} for ten main-sequence {\emph{Kepler}} F-stars. The stars were chosen with at least ten observational mode frequencies per angular degree for $l=0,1,2$. Moreover this sample of ten stars covers, within the constraints above, the larger possible range of values of large-frequency separations and effective temperatures. For a better visualisation of the results, the  stars of the sample  were separated into two groups according to the values of their observational effective temperatures (Table \ref{table:1}).

The characteristic quasi-periodic oscillation of the seismic parameter $\beta(\nu)$ is induced by the abrupt variation of the sound speed in the relevant ionisation regions \citep[e.g.,][]{2001MNRAS.322..473L}. We can verify from Figure \ref{fig1} that in hotter stars the amplitudes of the quasi-periodic oscillations are larger than in the case of the cooler stars. The seismic parameter $\beta(\nu)$ exposes the distinct behaviours/signatures of partial ionisation processes for cool and for hot main-sequence F-stars. For quantifying the different ionisation patterns among these stars, \citet{2017ApJ...843...75B} introduced two ionisation indexes $\Delta \beta_1$ and $\Delta \beta_2$. These indexes are defined as the amplitudes of the local maximums around the value of the frequency of the maximum power as illustrated in Figure \ref{fig2}. They serve as indicators of the magnitude of ionisation processes occurring in the outer layers. Plotting these indexes against the surface rotation rates of the stars (Figure \ref{fig3}) lead us to a relation in the form of a power law between the ionisation indexes and the rotation period of the stars \citep{2017ApJ...843...75B}. Also shown in Figure \ref{fig3} are the relationships between the ionisation indexes and other stellar fundamental observable properties, namely, the effective temperature and the metallicity. In both cases, the general trends are highlighted with a linear fit that supports the idea that ionisation is a key ingredient to understand the relationships between activity and rotation. 
The metallicity trend is also explicit in showing that high-metallicity stars correlate to higher values of the ionisation indexes, whereas low-metallicity stars correlate to lower values of the ionisation indexes. This is a very promising connection to further investigate the connection between metallicity and the dynamo properties of stars \citep[e.g.,][]{2018ApJ...852...46K}. Nevertheless, due to large error bars in metallicity measurements this trend should be considered with care.

\section{Partial ionisation zones in the outer convective envelopes of F-stars}\label{sec3}

\begin{figure} 
	\centering
	\includegraphics[width=0.45\textwidth]{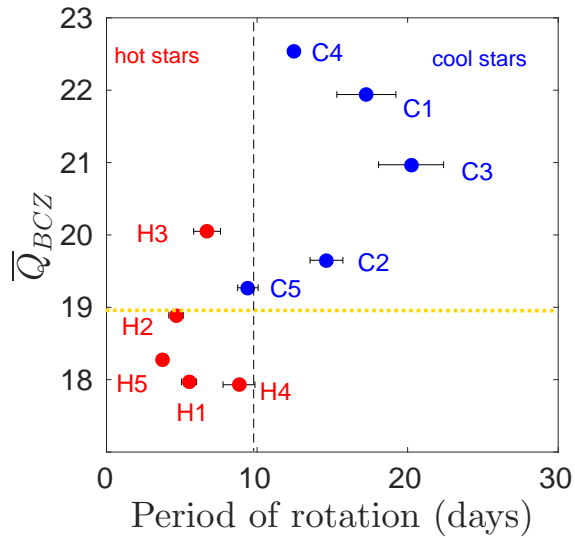}
	\caption{The mean effective ionic charge (Equation \ref{eq1}) taken at the BCZ for each stellar model and plotted against the surface rotation periods of stars. The vertical dashed line indicates the value of the rotation period around which occurs the transition in the rotational regimes. The orange horizontal  dashed lines indicates the beginning of K-shell ionisations for the heavy elements considered in the stellar models.}
	\label{fig10}
\end{figure}

\subsection{The mean effective ionic charge}

	Stellar matter in main-sequence F-stars is mostly composed of ions and free electrons. In stellar interiors each atomic species is, generally, in different stages of ionisation in consonance with the depth inside the star. The first element to reach complete ionisation is hydrogen when the temperature attains approximately $10^4$ K. Then, helium is the next element to reach full ionisation around $10^5$ K. At this temperature heavy elements like, for instance, carbon, nitrogen or oxygen, also have already lost some of their valence electrons. Gradually, with deepening, i.e., towards the centre of the star, atomic species heavier than hydrogen and helium increasingly lose electrons. Some of them become fully ionised. The Saha ionisation equation \citep{1921RSPSA..99..135S} gives the relative concentrations of atoms in consecutive degrees of ionisation. Therefore, it allows us to compute, for each chemical species, the ionisation fractions, as well as, the mean ionic charges of the different atomic species. With the purpose of  characterising the ionisation patterns in this sample of ten F-stars we define a mean effective ionic charge
	\begin{equation}\label{eq4}
	\overline{Q}= \langle Q_{\text{H}} \rangle + \langle Q_{\text{He}} \rangle + \langle Q_{\text{C}} \rangle + \langle Q_{\text{N}} \rangle + \langle Q_{\text{O}} \rangle  ,
	\end{equation}	
	where
	\begin{equation}\label{eq5}
	\langle Q_{\text{elem}} \rangle \equiv \sum_{j=1}^{q_{\text{elem}}} j \, x_{j, \text{elem}} \, \,   ,
	\end{equation}
	is the mean ionic charge of each element.
	Here, $q_{\text{elem}}$ is the atomic number of the considered atomic species and $x_{j, \text{elem}}$ represents the ionisation fraction of the atomic species in the ionisation stage $j$. 
	The ionisation stage of an atom is related to the atom's charge indicating the number of electrons missing. By ionisation fraction, $x_{j,\text{elem}}$, we understand the number of atoms of some element in the ionisation stage "j" divided by the total number of atoms of the mentioned element. Each one of the ionisation fractions is obtained numerically from a set of Saha equations as defined in \citep[][chap.~14]{2012sse..book.....K}. Therefore, the mean effective ionic charge defined by equation \ref{eq4}, contains information about the chemical abundances of the elements considered, giving the possibility of comparing radial profiles of different ionic charges among stars with different metallicities. Moreover, this charge focus on the positively charged ions allowing us to identify with precision the locations of the ionisation regions for all the elements included in theoretical stellar models. Free electrons from one element, as it is well known, can recombine with different ions of other elements influencing the overall equilibrium of the ionisation processes.
	
	The mean effective ionic charge in this work takes into account the mean ionic charges of five chemical elements: hydrogen, helium, carbon, nitrogen and oxygen. This particular choice of heavy elements is motivated by the fact that they are three of the most abundant elements in the Sun \citep[e.g.,][]{2009ARA&A..47..481A, 2011RPPh...74h6901T}. Until now there are no reasons to think that, in general, other main-sequence F-stars would have different chemical abundances \citep{2010Ap&SS.328..185G}. Moreover, concerning the Sun, these elements are known to be particularly important to understand the impact of the chemical composition of the deep convective zone on the sound speed \citep[e.g.,][]{2012ApJ...745...10B, 2016JPhCS.665a2078T}.

\subsection{Two Relevant Ionisation Zones: The Location of The Base of The Convective Zone}

To compute the mean ionic charges we have modelled our set of stars with the CESAM stellar evolution code \citep{1997A&AS..124..597M}.
The input physics of the models is as follows: OPAL equation of state \citep{2002ApJ...576.1064R} as well as OPAL opacities, nuclear reaction rates from the NACRE compilation \citep{1999NuPhA.656....3A} and the new abundances of \citet{2009ARA&A..47..481A}. Microscopic diffusion was taken into account using the Burgers formalism \citep{1969fecg.book.....B}. The atmosphere follows an Eddington-grey law and convection is treated according to the \citet{1958ZA.....46..108B} mixing-length theory without overshoot.
Table \ref{table:2} summarises the resulting properties of the theoretical stellar models.

To better understand the impact of the variations of the mean ionic charge we describe with some detail the partial ionisation processes in two specific stars (Figures \ref{fig4} and \ref{fig5}).
One is a cool star (C1), a representative of the slow rotators and the other a rapidly rotating hot star (H5). Their effective temperatures and rotation periods are given respectively by the pairs ($6033 \pm 77$ K, $17.26 \pm 1.26$ days) and ($6331 \pm 77$ K, $3.70 \pm 0.35$ days). Effectives temperatures were taken from \citet{2017ApJ...835..173S} and rotation periods from \citet{2015MNRAS.452.2654B}. The top panel of Figures \ref{fig4} and \ref{fig5} shows all the ionisation fractions for the five chemical species studied. Superimposed with the ionisation fractions we plotted the normalised mean effective ionic charge, $\overline{Q}_N=\overline{Q}/\max{\overline{Q}}$.  
For defining the boundaries of the ionisation regions we consider a definite range of temperatures as indicated in Figure \ref{fig4}. Ionisation region 1 (IR1) spreads over an interval from $4\times10^4$ K to $2\times10^5$ K. The relevant partial ionisation taking place in this region is assigned to helium (the two helium ionisations), since here hydrogen is already fully ionised. Nevertheless, we note that in this region also occurs the loss of most of the valence electrons from carbon, nitrogen and oxygen. The ionisation region 2 (IR2) is exclusively related to partial ionisation of heavy elements. It spreads over temperatures from $4\times10^5$ K to $1\times10^6$ K. For this temperature range, light elements are fully ionised and all the relevant ionisation processes are ruled by heavy elements. Particularly important to the variation of the ionic charge is the loss of the last electron of the L-shell, and also the K-shell ionisations in the three considered elements (CNO). The bottom panel of Figures \ref{fig4} and \ref{fig5} represents the gradient of the mean effective ionic charge with respect to the acoustic depth, $\tau = \int_{r}^{R}\frac{dr}{c_s}$, with  $c_s$ being the adiabatic sound speed, i.e., 
\begin{equation}\label{eq6}
	\nabla \overline{Q} \equiv \frac{d\overline{Q}}{d\tau}.
\end{equation}
This gradient allows us to spot the significant temperature intervals from the point of view of the variations of the ionic charges.

Comparing the partial ionisation profiles between the two stars, C1 and H5, we can notice that the main differences are related to the partial ionisation region of heavy elements IR2. In the hotter star, H5, IR2 extends itself over a larger radial section of the star. This fact is especially important since we will include in our analysis the locations of the base of the convective zone in each case, or more precisely, the relative locations of the BCZ and IR2. In the case of the cool star, C1, the BCZ is located outside IR2 (Figures \ref{fig4}, \ref{fig5}), whereas in the case of the hot star the BCZ is located inside the region IR2 (Figures \ref{fig4}, \ref{fig5}). From another perspective, we can say that most of the partial ionisation zone of heavy elements lies above the BCZ for the cool star, whereas for the hot star the partial ionisation region of heavy elements lies mostly below the BCZ. 
For completeness, in Figures  \ref{fig4} and \ref{fig5} are shown the theoretical and observational locations of the BCZ. The theoretical values are obtained directly from stellar models and are given in Table \ref{table:2}. The observational values for the location of the BCZ represented in Figures  \ref{fig4} and \ref{fig5} were measured seismically by \citet{2017ApJ...837...47V}. We see that in the case of the cool star, C1, a sun-like star, there is a very good agreement between the theoretical and observational values. In the case of the hot star, H5, the values do not coincide, but nevertheless, the disagreement is not large. This disagreement might be related with the description of the microphysics (e.g., equation of state, opacities) of the stellar interior in the case of the hot stars \citep[e.g.,][]{2018ApJ...853..183B}. As it is well known, the location of the BCZ is extremely sensitive to opacities \citep[e.g.,][]{2016JPhCS.665a2078T}.

Figure \ref{fig6} shows the evolution of the location of the BCZ during the pre-main sequence stage. Concerning the cool star, the BCZ will not cross the region IR2 during the pre-main sequence evolution, and when the star reaches the main sequence its BCZ will be located below the IR2. In contrast, for a hotter star, the BCZ will cross the region IR2 as a result of the combination of two factors: the larger radial extension of IR2 and a thiner convective envelope in hot stars. When the hot star reaches the main sequence, the BCZ will be located inside IR2. As we will discuss below, this structural difference linked to the relative locations of the ionisation region IR2 and the BCZ seems to be related to the different rotational regimes and magnetic properties of this sample of F-stars.

It is possible that the values and variations of the mean effective ionic charge have influence in the type of dynamo regime taking place in the star since it seems two regimes can be related to distinct values of the magnetic Prandtl number \citep{2019arXiv190400225C}. The magnetic Prandtl number can be defined by the \citet{1965RvPP....1..205B} plasma atomic diffusivities, which in turn, depend on the ionic charges. Another possibility is that these values are important for the redistribution of angular momentum mechanisms, such as, the Tayler-Spruit dynamo through the compositional component of the Brunt-V\"{a}is\"{a}l\"{a} frequency \citep[e.g.,][]{2019MNRAS.485.3661F}. Also, a battery-like mechanism \citep[e.g.,][]{2017A&A...604A..66K}, acting as an additional source of magnetic field at the base of the convective zone would benefit from variations of ionic charges. All these hypotheses require further investigation since, on the other hand, and concerning free electrons, the contribution of heavy elements accounts for approximately half free electron per atom.

\subsection{The Mean Effective Ionic Charges and their relation with stellar parameters}

Figure \ref{fig7} shows the mean effective ionic charges (Equation \ref{eq4}) plotted against the acoustic depth for each stellar model. 
We note that the acoustic extension of the ionisation region of light elements (IR1) varies less than the acoustic extension of the ionisation region of heavy elements (IR2). Moreover, the ionisation region IR2 is always larger than the ionisation region IR1. This effect is exacerbated in the case of hot stars (Figure \ref{fig7}).  In a rigorous representation, such as the one done in Figures  \ref{fig4} and \ref{fig5} for two specific cases, each star has its own boundaries between the two regions defined in a fixed temperature range.

The rate of change of the mean ionic charge in each region is defined by the slopes of the associated quasi-linear functions in each region ($s_1$ and $s_2$). These slopes can also be used as indicators of the magnitude of the partial ionisation processes occurring in stellar interiors. Figure \ref{fig8} represents the slopes computed for the two distinct regions (region IR1 and region IR2) and plotted against different stellar parameters. This figure shows how these two quantities are correlated with all the fundamental stellar properties represented (surface rotation period, effective temperature, metallicity, age, and mass). 
High values of the slopes are linked to cool stars whereas low values of the slopes are linked to hot stars (the star H4 is an outlier relatively to the variation rate in the ionisation zone of light elements). 

\section{The rotational regimes: a correlation with the location of the partial ionisation zone of heavy elements}\label{sec4}

Here, we investigate how the structural differences concerning the characteristics of the partial ionisation zones can be related to the rotational profiles of two studied groups of stars.
We have just seen that during the main-sequence, the base of the convective envelope for a cool star will be located predominantly below the partial ionisation region of heavy elements. In contrast, for a hotter star, the base of the convective envelope will be located predominantly above the partial ionisation region of heavy elements. We intent to quantify these structural differences by calculating the relative distance between the location of the BCZ and the region IR2. This relative distance, which we represent by $d_2$, is actually a quantity that represents the difference between the acoustic depths of the two locations: BCZ and IR2. We define it as,
\begin{equation}\label{eq7}
	d_2=\tau_{\text{IR2}}-\tau_{\text{BCZ}}\, ,
\end{equation}
where $\tau_{\text{IR2}}$ is the acoustic depth of IR2 (taken at its central value) and $\tau_{\text{BCZ}}$ is the acoustic depth of the BCZ. The acoustic depth corresponding to the location of the base of the convective envelope can be obtained from the theoretical models (Table \ref{table:2}) and from observational data using asteroseismology \citep{2017ApJ...837...47V}.

From the acoustic distance defined by Equation \ref{eq7}, three possibilities emerge. The first possibility is the case in which $\tau_{\text{BCZ}}>\tau_{\text{IR2}}$, resulting in a negative value for $d_2$.  This means that the BCZ is predominantly below the IR2, which in turn, represents the typical structure of the cooler stars. The second possibility occurs when $\tau_{\text{BCZ}}<\tau_{\text{IR2}}$, leading to a positive value for $d_2$. Consequently, the BCZ is predominantly above the IR2 and this corresponds to the typical structure of the hotter stars. The third possibility arrives when $\tau_{\text{BCZ}}\simeq\tau_2$. In this situation the BCZ is located in the central region of IR2. This condition, by definition, represents a transition in the structure of the star, a transition between the two cases discussed above. The top panel of Figure \ref{fig9}, shows the rotation periods of all the ten studied stars plotted against the distance $d_2$. It allows us to relate the above mentioned third possibility to the transition between the distinct rotational regimes in each group of stars (cool and hot stars). We can see that as the distance $d_2$ approaches zero, the rotation period approaches the critical value of 10 days ($d_2 \rightarrow 0 \Rightarrow P_{\text{rot}} \rightarrow 10$ days). This trend is compatible with the transition in the rotational regimes of main-sequence F-stars, usually known as the Kraft break. We can also identify large positive values of $d_2$ with short rotation periods and large negative values of $d_2$ with longer rotation periods. These possibilities are illustrated schematically in Table \ref{table:3}.
It seems that the acoustic distance $d_2$ noticeably separates the two rotational regimes for F-type stars.
For testing the consistency of this result we have also plotted the effective temperatures against the acoustic distance $d_2$. As the distance $d_2$ approaches zero, the effective temperature approaches the critical value of 6250 K ($d_2 \rightarrow 0 \Rightarrow T_{\text{eff}} \rightarrow 6250$ K). 
Finally, we represented the photometric magnetic activity index $\langle S_{\text{ph}} \rangle$ \citep{2014A&A...562A.124M} against $d_2$ (bottom panel of Figure \ref{fig9}). The values for the photometric index, $S_{\text{ph}}$, were taken from \citet{2014A&A...572A..34G}. In this case the trends are not so clear but still it is possible to relate the more active stars with a positive value of $d_2$, i.e., with stars where the region IR2 is located predominantly below the BCZ.

The results expressed in Figure \ref{fig9} unveil a possible relation  between the two structural types of main-sequence F-stars (Figures \ref{fig4} and \ref{fig5}) and the two main rotational regimes observed for these stars \citep[e.g.,][]{1967ApJ...150..551K}. The two types of different stellar structures are seemingly related to the microphysics of the star, more specifically, to partial ionisation processes occurring in the interiors of these stars.

Finally, Figure \ref{fig10} shows a relationship between the value of the mean effective ionic charge (Equation \ref{eq4}) taken at the base of the convective envelope for each stellar model ($\overline{Q}_{\text{BCZ}}$) and the surface rotation periods of stars. We found that, for this sample, the cooler stars admit values of the mean effective ionic charge $\overline{Q}_{\text{BCZ}} \gtrsim 19$, whereas hotter stars are more likely to have  $\overline{Q}_{\text{BCZ}} \lesssim 19$. The meaning of $\overline{Q}_{\text{BCZ}} =19$ is interesting from the viewpoint of atomic physics. Taking into account the chemical composition considered in Equation \ref{eq4}, $\overline{Q}_{\text{BCZ}} =19$ indicates the beginning of the K-shell ionisations in heavy elements. K-shell ionisation are important contributors to the variation of $\overline{Q}$. The first K-shell ionisation of carbon is localised approximately in the central region of the ionisation region IR2. Hence the choice of this central region to the definition of the distance $d_2$ (Equation \ref{eq7}).

\begin{table}
	\centering
	\caption{Distinct stellar structures according to $d_2$}
	\begin{tabular}{  | l | l | }
		\hline
		& $< 0$ structure of a slowly rotating cool F-star  \\
		$d_2$ & $\approx 0$ transitional case  \\
		&  $> 0$ structure of a rapidly rotating hot F-star \\
		\hline
		\label{table:3}
	\end{tabular}
\end{table}

\section{Conclusions}\label{sec5}

It is well known that late and early-type F-stars have different rotational rotational behaviours on the main sequence. The former spin down with time whereas the latter keep the rapid rotation profile from the pre-main-sequence phase of evolution.
We found that a very particular structural characteristic of the studied sample of F-stars, i.e., the distance between the partial ionisation region of heavy elements (carbon, nitrogen and oxygen) and the base of the convective envelope correlates with the transition between rapid and slow rotators.

From a physical standpoint, it is important to highlight that regions of partial ionisation of chemical elements are regions of large variations of ionic charges and that electric charges are among the main ingredients of the generation of magnetic fields.
It is known that the effect of internal magnetic fields is tightly related to the redistribution of angular momentum in the stellar interiors, which in turn, cannot be separated from the internal chemical element transport \citep[e.g.,][]{2017RSOS....470192S}. Among the physical processes that act to redistribute angular momentum throughout the interior of the star are several hydrodynamical instabilities \citep[e.g.,][]{2013EAS....62..227P}, such as, for example, the Tayler-Spruit magneto-rotational instability \citep{1973MNRAS.161..365T, 2002A&A...381..923S}. A recent work that uses a modified prescription of this mechanism \citep{2019MNRAS.485.3661F} brought the core rotation rates for stars in different evolutionary stages in good agreement with data from asteroseismolgy. The surface rotation rates of solar-type stars were also subject of studies involving the Tayler-Spruit dynamo \citep[e.g.,][]{2007ApJ...655.1157D, 2010ApJ...716.1269D}, where it was shown that the Tayler-Spruit dynamo cannot explain the observed values for the rotation periods of the studied solar-type stars. It is important though to highlight that the different Tayler-Spruit redistribution mechanisms by magnetic torques strongly rely on the compositional component of the Brunt-V\"{a}is\"{a}l\"{a} frequency, $N_\mu$, which in turn relies on the concept of mean molecular weight. The mean molecular weight can be defined to take into account the characteristics of a partially ionised gas \citep[e.g.,][]{2012sse..book.....K}. Although in the central part of the star partial ionisation might be considered not very important, in the more external layers partial ionisation plays an important role. Our study indicates that taking into account partial ionisation can contribute to improve our understanding of magneto-hydrodynamical instabilities like the Tayler-Spruit dynamo.
		
Still concerning the impact that magnetic fields might have on the redistribution of internal angular momentum, it is worth to remember that magnetic fields are thought to be generated in stellar interiors by convective dynamos that amplify a seed remnant field from the star's formation process. Another recent work considers the characteristics of two distinct convective dynamo regimes that can be associated to low-mass stars versus high-mass stars \citep{2019arXiv190400225C}. In these two distinct regimes the magnetic Prandtl number, $P_m=\nu/\eta$, defined as the ratio of kinematic viscosity to magnetic diffusity plays a predominant role. The two regimes also seem strongly related to the location of the convective zones. The magnetic Prandtl number can be defined using the Braginskii atomic diffusivities \citep{1965RvPP....1..205B} yielding an explicit dependence on the mean ionic charge. This dependence of the magnetic diffusivity and kinematic viscosity on the mean effective ionic charge was used by \citet{2019arXiv190400225C} in their study of Rossby and Prandtl number scaling relationships for stellar dynamos.
		 
Moreover, activity-rotation relations \citep[e.g.,][]{1984ApJ...287..769N, 2003A&A...397..147P} based on the stellar Rossby number, which corresponds to the ratio of the rotation period to the convective overturn time ($R_{\text{o}s} = P_{\text{rot}}/\tau_\text{c}$), suggest a link between rotation and the efficiency of the dynamo mechanism in the stellar interiors. The stellar Rossby number is just one, among the several possible definitions for the Rossby number \citep[e.g.,][]{2017ApJ...836..192B}, though it is not completely clear how these definitions express the complex mutual impact of rotation and convection \citep{2019ApJ...872..138A}. Being the Rossby number a locally defined quantity, the knowledge that the locations of partial ionisation zones can be important to characterise the rotational properties of F-stars might translate into an improvement in the calculation of the Rossby number, and in a better understanding of the dynamo in general.

The path to the understanding of the different stellar rotational behaviours also implies learning about the mechanisms responsible for the primordial origin of  magnetic fields. The primordial origin of magnetic fields is a major unresolved astrophysical problem \citep[e.g.,][]{2008RPPh...71d6901K, 2015SSRv..191...77F}. A relic magnetic field is usually assumed to exist since the star-formation time in the Sun and other main-sequence stars with external convective envelopes \citep[e.g.,][]{2017LRSP...14....4B, 2010LRSP....7....3C, 2013SAAS...39.....C, 2014ARA&A..52..251C}. This magnetic field is then amplified and sustained by the action of a dynamo mechanism taking place at the base of the convective envelope, where the differential rotation plays a crucial role \citep[e.g.,][]{2013ApJ...777..153A, 2017LRSP...14....4B}. Nevertheless, one possibility of spontaneously creating a seed magnetic field in stars was proposed by Biermann \citep{1950ZNatA...5...65B}, and is usually known as the Biermann mechanism \citep[e.g.,][]{1979ApJ...229.1126L}. From a physical point of view the Biermann mechanism is based on the difference between the values of the masses of ions and electrons. When in the presence of a centrifugal field, ions and electrons will move differently originating a flow of electron currents relative to ions \citep{1951PhRv...82..863B}. 
The effect of the Biermann battery is usually not considered because it is assumed that the corresponding terms in the magnetic's field induction equation are relatively small \citep[e.g.,][]{1962ApJ...136..615M}. Nevertheless, recent numerical simulations for the Sun show that the action of a Biermann battery mechanism can generate fields of order of $10^{-6}-10^{-2}$ Gauss, which together with dynamo amplification, are  able to explain solar observations for the quiet Sun \citep{2017A&A...604A..66K}.

The theory of angular momentum evolution in stars is complex and challenging because it should take into account the origin of stellar rotation, the transport and redistribution of angular momentum in stellar interiors and the loss of angular momentum through magnetised stellar winds. Possibly, for better understanding angular momentum transport in F-stars we will need to consider more than one mechanism. Furhter theoretical studies that take into account the details of the microphysics of stellar interiors are needed to better explain the always increasing amount of observational data.

Other relevant and directly related fields that might benefit from this result are gyrochronology and exoplanetary astrophysics. Gyrochronology intents to establish a mass-age-rotation relation \citep[e.g.,][]{1972ApJ...171..565S, 2007ApJ...669.1167B,  2008ApJ...687.1264M, 2013ApJ...776...67V, 2015Natur.517..589M}. This is a promising and important approach to know the ages of stars. Stellar ages are among the most difficult parameters to determine with precision. Still, here too, some problems persist. There is currently an upper mass limit for the rotation-age relationships to be valid. This limit coincides with the transition between rapid and slow rotators ($\sim 1.2 \, M_\odot$). Gyrochronologic relations are confined to the $0.6-1.2 \, M_\odot$ domain, and even in this domain were found discrepancies with ages calculated using asteroseismology \citep[e.g.,][]{2015MNRAS.450.1787A}. Concerning exoplanetary astrophysics, it is known that many of the already discovered planet hosts are sun-like stars. These, in turn, have convective outer envelopes with different depths. It is striking that among the host stars, rapid rotators tend to exhibit a dearth of planets orbiting around them. In contrast, most known host stars are slow rotators \citep[e.g.,][]{2017arXiv170804449M}. 
Learning about rotation rates of stars and its relation to the microphysics of stellar interiors might greatly contribute to improve our knowledge about different and influential fields in stellar astrophysics.

\section*{Acknowledgements}

The authors would like to thank the anonymous referee for all the remarks, comments, and sugestions that improved the quality of this manuscript.
We are also grateful to P. Morel for making available the CESAM code for stellar evolution and 
to Jordi Casanellas for the modified version of the same code and for his valuable help. This research has made use of the Aarhus adiabatic pulsation code (ADIPLS) and we acknowledge his author J. Christensen-Dalsgaard.
The authors thank the Funda\c c\~ao para a Ci\^encia e Tecnologia (FCT), Portugal, for the financial support to the Multidisciplinary Center for Astrophysics (CENTRA),  Instituto Superior T\'ecnico,  Universidade de Lisboa,  through the Grant No. UID/FIS/00099/2013.
This work was also supported by grants from "Funda\c c\~ao para a Ci\^encia e Tecnologia" (SFRH/BD/74463/2010).




\bibliographystyle{mnras}
\bibliography{biblio} 




%
%


\bsp	
\label{lastpage}
\end{document}